\documentclass[fleqn,usenatbib]{mnras}
\usepackage{newtxtext,newtxmath}

\usepackage[T1]{fontenc}

\DeclareRobustCommand{\VAN}[3]{#2}
\let\VANthebibliography\thebibliography
\def\thebibliography{\DeclareRobustCommand{\VAN}[3]{##3}\VANthebibliography}

\usepackage{graphicx}	
\usepackage{amsmath}	
\usepackage[section]{placeins}

\title[The radio continuum emission of {\sc Shark} galaxies]{Modelling the galaxy radio continuum from star formation and active galactic nuclei in the {\sc Shark} semi-analytic model}

\author[S. ~P. Hansen et al.]{
\parbox[t]{\textwidth}{
\vspace{-0.5cm}
Samuel P. Hansen$^{1}$\thanks{E-mail: samuel.hansen@icrar.org} Claudia D.~P. Lagos$^{1,2,3}$, Matteo Bonato$^{4,5}$, Robin H. W. Cook$^{1}$ Luke J. M. Davies$^{1}$, Ivan Delvecchio$^{6}$, Scott A. Tompkins$^{1,7}$}
\vspace*{6pt} \\
$^{1}$International Centre for Radio Astronomy Research (ICRAR), M468, University of Western Australia, 35 Stirling Hwy, Crawley, \\WA 6009, Australia.\\
$^{2}$ARC Centre of Excellence for All Sky Astrophysics in 3 Dimensions (ASTRO 3D).\\
$^{3}$Cosmic Dawn Center (DAWN), Denmark.\\
$^{4}$INAF-Istituto di Radioastronomia, Via Gobetti 101, 40129 Bologna, Italy. \\
$^{5}$Italian ALMA Regional Centre, Via Gobetti 101, 40129 Bologna, Italy. \\
$^{6}$INAF - Osservatorio Astronomico di Brera, via Brera 28, I-20121, Milano, Italy \& via Bianchi 46, I-23807, Merate, Italy.\\
$^{7}$ School of Earth and Space Exploration, Arizona State University, Tempe, AZ 85287-1404.}
\date{Accepted 2024 May 06. Received 2024 April 30; in original form 2023 October 19}
\pubyear{2024}

\begin{document}
\label{firstpage}
\pagerange{\pageref{firstpage}--\pageref{lastpage}}
\maketitle

\begin{abstract}
We present a model of radio continuum emission associated with star formation (SF) and active galactic nuclei (AGN) implemented in the \textsc{Shark} semi-analytic model of galaxy formation. SF emission includes free-free and synchrotron emission, which depend
on the free-electron density and the rate of core-collapse supernovae with a minor contribution from supernova remnants, respectively. AGN emission is modelled based on the jet production rate, which depends on the black hole mass, accretion rate and spin, and includes synchrotron self-absorption. \textsc{Shark} reproduces radio luminosity functions (RLFs) at $1.4\,\rm GHz$ and $150\,\rm MHz$ for $0\le z \le 4$, and scaling relations between radio luminosity, star formation rate and infrared luminosity of galaxies in the local and distant universe in good agreement with observations. The model also reproduces observed number counts of radio sources from $150$~MHz to $8.4$~GHz to within a factor of two on average, though larger discrepancies are seen at the very bright fluxes at higher frequencies. We use this model to understand how the radio continuum emission from radio-quiet AGNs can affect the measured RLFs of galaxies. We find current methods to exclude AGNs from observational samples result in large fractions of radio-quiet AGNs contaminating the “star-forming galaxies” selection and a brighter end to the resulting RLFs.We investigate how this effects the infrared-radio correlation (IRRC) and show that AGN contamination can lead to evolution of the IRRC with redshift. Without this contamination our model predicts a redshift- and stellar mass-independent IRRC, except at the dwarf-galaxy regime. 

\end{abstract}



\section{Introduction}\label{sec_introduction}

The radio sky provides an excellent laboratory for studying galaxy populations. Galaxy radio emission is understood to arise from two main processes; star-formation (SF) and active galactic nuclei (AGN). The link between radio emission and SF arises through synchrotron radiation produced by the acceleration of cosmic ray electrons by the magnetic fields associated with core-collapse supernovae (CCSNe) and free-free emission from the interactions of free electrons in HII regions around young, massive stars (\citealt{condon1992radio}). Because radio emission is not affected by dust, it is thought to be an excellent tracer of the  star formation rate (SFR) in galaxies.

The radio continuum emission associated with AGNs is also due to synchrotron radiation but this time powered by collimated jets of ejected plasma (\citealt{panessa2019origin}) in the vicinity of super massive black holes (SMBHs) at the centre of galaxies. AGNs that predominantly emit in the radio continuum are termed ``radio-loud'' and are generally bright, $\rm S \gtrsim 1 mJy$ around 1~GHz (\citealt{padovani2017active}). Fainter radio AGNs are termed ``radio-quiet'' and, while they can still have jets, most of their electro-magnetic output happens in other wavelengths. At $\rm S \lesssim 1 mJy$ around 1~GHz thus, observations get a mix of emission associated with SF and radio-quiet AGNs.

Radio continuum emission has been used extensively as a SFR tracer in observations, and this obstensively rests on the existence of 
the infrared-radio emission correlation (a.k.a. the Infrared Radio Correlation, IRRC; \citealt{bell2003estimating, condon2016essential, duncan2020mosdef, molnar2021non}). The IRRC is an observed tight correlation between a galaxy's  total IR luminosity and radio luminosity that spans five orders of magnitude  (\citealt{van1971observations,van1973high,de1985spiral,helou1985thermal,condon1992radio}).  It has been shown to exists in a variety of different galactic populations including Sub-Millimetre Galaxies (SMGs) (\citealt{thomson2014alma,algera2020alma}, (Ultra)-Luminous Infrared Galaxies [(U)LIRGs] (\citealt{lo2015combining}), Early-Type galaxies (\citealt{omar2018far}), dwarf galaxies (\citealt{shao2018local}), low-ionization nuclear emission-line region (LINERs) and Seyferts (\citealt{solarz2019radio}), irregular and disk-dominated galaxies (\citealt{pavlovic2021does}) as well as highly-magnified galaxies (\citealt{giulietti2022far}), to name a few. 

The IRRC is often parameterised by $q_{\rm IR}$ \citep{helou1985thermal} as

\begin{equation} \label{eqn_qir}
\mathrm{
    q_{\rm IR} = {\rm log}_{10} \left(\dfrac{L_{\rm IR}}{3.75\times 10^{12}\, \rm W\, m^{-2}}\right) - {\rm log}_{10} \left(\dfrac{L_{\rm rad,1.4GHz}}{\rm W\, m^{-2}\, Hz^{-1}} \right),
    }
\end{equation}

\noindent where $\rm L_{IR}$ is the total IR luminosity integrated over the wavelength range $\rm 8 - 1000 \mu m$ in the rest frame and $ L_{\rm rad,1.4GHz}$ the total rest-frame radio luminosity at $\rm 1.4 GHz$. 

The IRRC's importance to the broader understanding of the link between star-formation, radio and infrared (IR) emission makes it an active area of research. In particular, there is debate in the literature about any evolution of $q_{\rm IR}$ with redshift, with some finding $q_{\rm IR}$ decreases with increasing redshift (\citealt{ivison2010blast,ivison2010far,magnelli2015far,delhaize2017vla}) and others finding no evolution (\citealt{appleton2004far,jarvis2010herschel,sargent2010vla,sargent2010no,bourne2011evolution,mao2011no,duncan2020mosdef,thomson2014alma,algera2020alma,cook2024devils}). 

Suggested reasons for this discrepancy in the evolution of $q_{\rm IR}$  with redshift are varied and include possible biases in the galaxy populations being studied (\citealt{sargent2010no}) and biases arising from low number statistics (\citealt{jarvis2010herschel}). Further contamination by AGNs at high redshifts could also explain this trend (\citealt{delvecchio2021infrared,kirkpatrick2013goods}). To place these results into context it is useful to explore them in physical models of galaxy formation, which attempt to predict the formation and evolution of galaxies together with the multi-wavelength emission.

However, most of the efforts to model galaxy emission in physical models of galaxy formation, such as semi-analytic models of galaxy formation (SAMs) and cosmological hydrodynamical simulations has been primarily focused on the wavelength range from the far-ultraviolet (FUV) to the far-infrared (FIR) due to the sheer volume of observations available at those wavelengths \citep{somerville2012galaxy,lacey2016unified,camps2016far,trayford2017optical,lagos2019far,Shen2022High}.
Models of the radio sky have been done mainly through empirical or semi-empirical models.

\citet{wilman2008semi} produced a semi-empirical model of the radio continuum sky by sampling observed radio luminosity functions (RLF). Thus, it depends very sensitively on the capability of observations to being able to distinguish radio emission coming from SF or AGNs, which is something that is especially hard at high redshift where little multi-wavelength information is available.

Another model of radio emission is the Tiered Radio Extragalactic Continuum Simulation (T-RECS) (\citealt{bonaldi2019tiered}). This model is empirically based and models SFGs and AGNs separately from each other. Radio emission from SFGs are modelled using the calibration between free-free and synchrotron luminosities from \citet{murphy2011calibrating} and \citet{murphy2012star}. However, these calibrations were shown in \citet{mancuso2015predictions} to over-predict the faint-end of local RLFs compared with the results of \citet{mauch2007radio}. T-RECS adopts a synchrotron luminosity that artificially inflates the SFR of low-luminosity galaxies to amend this.

The next decade promises a suite of new telescopes and surveys looking further and deeper into the extragalactic radio sky, which makes it urgent to extend the predictive power of existing SAMs and cosmological hydrodynamical simulations to make predictions in this regime. Current surveys include the Low Frequency Array's (LOFAR) Two-metre Sky Survey (LoTSS), the Very Large Array Sky Survey (VLASS) and the GaLactic and Extragalactic
All-sky MWA (GLEAM)-X survey being undertaken at the Murchison Square Array (MWA). Such surveys have had recent data releases (\citealt{hurley2022galactic, shimwell2022lofar}) and have observations scheduled until as late as 2024 (\citealt{lacy2016vla}). These surveys also cover a wide range of radio frequencies from low frequency radio in LoTSS (120 - 168~MHz) (\citealt{hurley2022galactic,morganti2011lofar}) to high radio frequencies (2GHz - 4GHz) in VLASS (\citealt{lacy2016vla}). These surveys are all precursors to the largest radio telescope in the world - the Square Kilometre Array (SKA) which will be capable of observing distant objects with greater resolution than ever before (\citealt{jarvis2015star,bonaldi2019square}). 

With this in mind, in this paper we combine the \textsc{Shark} SAM \citep{lagos2018shark,lagos2023} with a theoretical model of radio emission from SF \citep{bressan2002far} (henceforth B02) and a theoretical model of radio emission from AGNs \citep{fanidakis2011grand} (henceforth F11) building on the work of \citet{lagos2019far}  to extend the predictive power of \textsc{Shark} by three orders of magnitude in frequency. This makes \textsc{Shark} the first SAM to directly model a large population of galaxies over cosmic time that include predictions of IR emission, radio emission and consequently $q_{\rm IR}$. 

This paper is organised as follows. In Section \ref{cpt_method} we introduce the model of radio emission and summarise how the FUV-to-FIR emission is modelled in \textsc{Shark}; Section~\ref{chp_results} presents key results from \textsc{Shark} and this model of radio emission, including (i) radio source counts across seven different radio frequencies; (ii) RLFs at 1.4~GHz and 150~MHz over a redshift range $\rm z = 0 - 4$; (iii) the properties of local SFGs; (iv) the properties of LIRGs and ULIRGs at high redshift; and (v) the evolution of $q_{\rm IR}$ with redshift and its dependence on stellar mass ($M_{\rm *}$). These are all compared with relevant observational results to test the capabilities of the model. Section~\ref{sec_conclusion} summarises the main findings of this work and presents our conclusions.

\section{Methods}\label{cpt_method}

This section introduces the methods used throughout this paper and how radio emission due to SF and AGNs is modelled in \textsc{Shark}. Fig. \ref{fig_ir_rad_emission_model} gives a brief visual representation of how these processes are modelled in each galaxy in \textsc{Shark}.

\subsection{The {\sc Shark} semi-analytic model of galaxy formation} \label{sec_shark}

\textsc{Shark} is an open source SAM of galaxy formation first presented in \citet{lagos2018shark} and recently updated in \citet{lagos2023}. The model includes physical processes that are thought to play an important role in shaping galaxy formation. These are (i) the merging and collapse of dark matter (DM) haloes; (ii) the accretion of gas on to haloes, which is controlled by the DM accretion rate; (iii) the shock heating and radiative cooling of gas inside DM haloes, leading to the formation of galactic disks via the conservation of specific angular momentum of the cooling gas; (iv) star formation in galaxy disks; (v) stellar feedback from the evolving stellar populations; (vi) chemical enrichment of stars and gas; (vii) the growth via gas accretion and merging of supermassive black holes; (viii) heating by AGNs; (ix) photoionisation of the intergalactic medium; (x) galaxy mergers driven by dynamical friction within common DM haloes, which can trigger starbursts and the formation and/or growth of spheroids; (xi) collapse of globally unstable disks that also lead to starbursts and the formation and/or growth of bulges. \textsc{Shark} v1.1 is adopted here as presented in \citet{lagos2018shark}. This is the same model adopted in modelling the UV-IR emission in \citet{lagos2019far}. \textsc{Shark} adopts a universal \citet{chabrier2003galactic} initial mass function (IMF).

The backbone of \textsc{Shark} is the Synthetic UniveRses For Surveys ({\sc SURFS}) simulation suite (\citealt{elahi2018surfs}), which is a set of N-body DM only simulations. The L210N1536 is what we use here, which has a boxsize of $\rm L_{box} = 210 cMpc/h$ and a softening length $\rm \epsilon = 4.5 ckpc/h$. Note that here cMpc and ckpc refer to comoving mega parsec and kilo parsec, respectively.
L210N1536 contains $\rm N_{p} = 1536^{3}$ DM particles each with a mass of $m_{\rm p} = 2.21 \times 10^{8} \rm M_{\odot}/h$. The 
simulation  adopts a $\rm \Lambda CDM$ cosmology with  a Hubble constant of $H_{\rm 0} = h \times 100\,\rm  (km/s)/cMpc$, $\rm h = 0.6751$, matter density of $\Omega_{\rm m} = 0.3121$, baryon density $\Omega_{\rm b} = 0.0491$ and dark energy density of $\Omega_{\rm \Lambda} = 0.6879$. 

This simulation produces 200 snapshots logarithmically arranged from $\rm z = 24 - 0$. This corresponds to a time between snapshots of $\rm \approx 6 - 80 Myr$. 

Halos, subhalos and their properties are identified using \textsc{VELOCIraptor} (\citealt{elahi2019hunting}, \citealt{canas2019introducing}). \textsc{VELOCIraptor} first identifies halos using a 3D friend-of-friend (FOF) algorithm. This 3D FOF structure corresponds to the halo.  It also applies a 6D FOF with velocity dispersion to remove spuriously linked objects (such as early stage mergers). It then identifies particles that have a local velocity distribution significantly different from the smooth background halo to identify substructure. It runs a phase-space FOF on these particles to identify the subhalos. {\sc SURFS} only considers halos with $\rm \geq 20$ DM particles.

Merger tress are then constructed using \textsc{TreeFrog} (\citealt{elahi2019climbing}). At its most basic, \textsc{TreeFrog} is a particle correlator that relies on particle IDs being continuous across snapshots. The merger tree is constructed forward in time, identifying the optimum link between progenitors and descendants. \textsc{TreeFrog} searches up to four snapshots to identify optimal links.

 \textsc{VELOCIraptor} and \textsc{TreeFrog} provide the subhalo and merger tree catalogues, respectively, which provide the basis from which galaxies are evolved. \textsc{Shark} evolves these galaxies across snapshots using a physical model. The physical model used here is fully described by Eqs. 49-64 in \citet{lagos2018shark}. Before this evolution takes place, the merger trees undergo a post processing treatment which is fully described in Section 4.1 of \citet{lagos2018shark}.

Within \textsc{Shark} there are three different types of galaxies; centrals, which are the central galaxy of the central subhalo; satellites, which are the central galaxy of satellite subhalos; and orphans, which are the central galaxy of a defunct subhalo. A defunct subhalo is one which has merged onto another subhalo and is not the main progenitor. From these definitions a central subhalo can have only one central galaxy, but many orphan galaxies, but a satellite subhalo can only have one galaxy. 

The key assumption of \textsc{Shark} and other SAMs is that galaxies can be fully described by a disk and a bulge. The fundamental difference between a galaxy's disk and bulge is the origin of the stars that constitute each component. Disk stars are formed from gas that is accreted onto the galaxy from the halo. Bulge stars can be either accreted from satellite galaxies that merge onto the central, or formed by a starburst episode driven by galaxy mergers or disk instabilities. The SFR within both components is driven by the surface density of molecular hydrogen, but this process is ten times more efficient in bulges. This higher efficiency was shown to be responsible for reproducing the observed cosmic star formation rate density for $\rm z \gtrsim 1.5$ (\citealt{lagos2018shark}). As described above, stars in the bulge can form due to disk instabilities when self-gravity dominates over centrifugal forces.

\subsection{Modelling Infrared Emission} \label{sec_IR_model}
\begin{figure*}
    \centering
    \includegraphics[scale = 0.4]{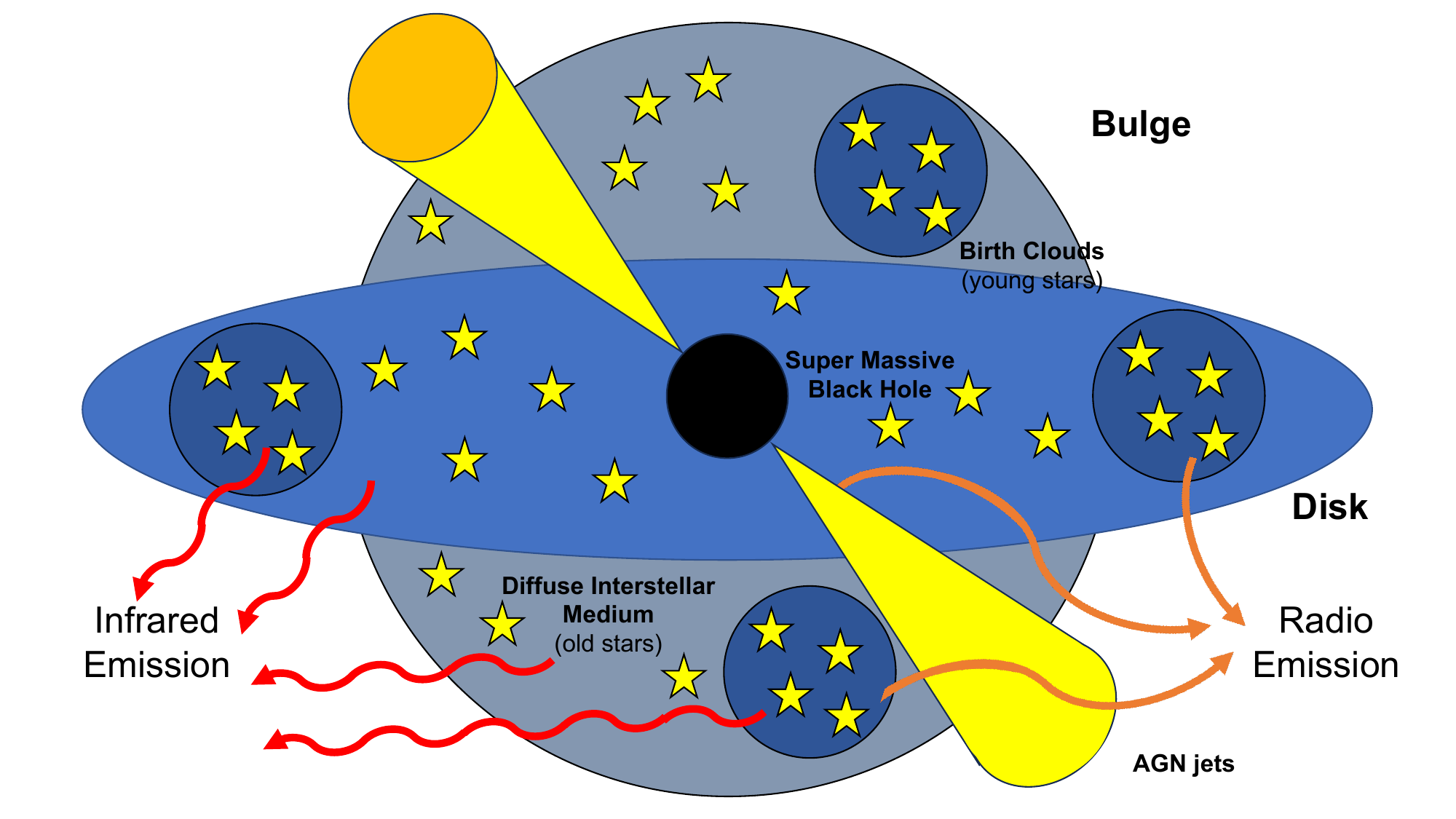}
    \caption{A diagram showing the physics behind the IR and radio emission. IR emission is modelled as in \citet{lagos2019far}. UV light from young stars is absorbed by dust and re-emitted at IR wavelengths. Dust is modelled as being in two phases: birth clouds (BC) and diffuse ISM Attenuation dust is modelled using the \citet{charlot2000simple} model, whose parameters depend on galaxy properties. UV light from young stars in BCs are attenuated by both the BCs and ISM, while older stars are only attenuated by the ISM.
    Radio emission is modelled as a combination of star-formation and AGNs contribution. Radio emission from SF is driven by the presence of young stars (those producing the bulk of the ionising radiation) and from supernovae, both of which are assumed to reside in BCs only. Radio emission by AGNs is driven by the power of jets being a combination of those in the Thin Disk (TD) and Advection Dominated Accretion Flow (ADAF) modes.}
    \label{fig_ir_rad_emission_model}
\end{figure*}

Fig.~\ref{fig_ir_rad_emission_model} shows a diagram of the physics behind the IR and radio emission in \textsc{Shark}. IR emission is modelled using the method presented in \citet{lagos2019far}. The simplest explanation is that IR is the result of UV light being attenuated by dust in birth clouds (BC) and the diffuse Interstellar Medium (ISM), and re-emitted in the IR. The UV light is attenuated using the \citet{charlot2000simple} model with attenuation parameters informed by the radiative transfer analysis of galaxies in the cosmological hydrodynamical simulations {\sc Eagle} presented in \citet{trayford2019fade}. This attenuated light is fully re-emitted in the IR following empirical IR templates from \citet{dale2001infrared}. We refer the reader to \citet{lagos2019far} for more details on how the FUV-to-FIR galaxy Spectral Energy Distributions (SEDs) are produced. 

{\sc Shark} uses the \textsc{ProSpect} \citep{robotham2020prospect} SED code in its generative mode to produce the synthetic SEDs. {\sc ProSpect} has the choice of also producing the radio continuum emission associated to star formation, by assuming an underlying IR luminosity to 1.4~GHz continuum luminosity ratio, the spectral indices of the synchrotron and free-free emission and the fraction of the radio continuum that is in the free-free component. This is what was used by \citet{tompkins2023cosmic} to make predictions of the galaxy number counts from $150$~MHz to $8.4$~GHz using {\sc Shark}. Note that this option in {\sc ProSpect} can only generate the radio continuum associated to star formation and not AGNs. However, assuming a constant $q_{\rm IR}$ is inflexible and limits the predictive power of {\sc Shark}. We circumvent this by including in {\sc Shark} two physical models to compute the radio continuum emission associated to star formation and AGNs, which we describe below.

\subsection{Modelling the Radio Emission associated with star formation} \label{sec_rad_method}

Radio emission in B02 is modelled as the sum of the free-free and synchrotron components from each galaxy. Free-free emission is based on the production of ionising photons (as the main source of free electrons) by young and massive stars. Synchrotron emission is modelled from CCSNe which accelerate cosmic ray electrons, making them emit photons at radio frequencies. Supernova remnants also make a minor contribution to synchrotron emission. Below we describe how we model both sources of radio continuum radiation.

\subsubsection{Free-Free Radiation}

B02 models free-free radiation to be proportional to the production rate of ionising radiation (i.e. Lyman continuum photons) as a proxy for the number of free electrons in the ISM.

The production rate of Lyman continuum photons ($\rm Q_H/[s^{-1}]$ )
is calculated as 

\begin{equation}
    Q_{\rm H} = \int_{0}^{\lambda_{0}} \left(\frac{\lambda  L_{\lambda}}{h\,c}\right) {\rm d}\lambda,
    \label{qh}
\end{equation}

\noindent where $\lambda_0$ is the Lyman limit, 921~\AA, $ \rm L_{\lambda}$ is the galaxy spectrum  in $\rm erg s^{-1}$ \AA $^{-1}$, here $h$ is Planck's constant (not to be confused with the Hubble parameter of Section \ref{sec_shark} ) and $c$ is the speed of light. $\rm L_{\lambda}$ is sourced from the galaxy spectrum created using the \textsc{ProSpect} {\it before} dust attenuation is applied.

In the process of photo-ionisation of hydrogen, Lyman continuum photons are absorbed. By assuming that the production rate is equal to the destruction rate of Lyman continuum photons, and building off the work of \citet{rubin1968discussion}, \citet{condon1992radio} expressed the free-free luminosity in terms of this production rate, HII regions gas temperature, $T$, and frequency, $\nu$,  as

\begin{equation}
    {\frac{L_{\rm ff}}{[\rm W \,Hz^{-1}]} = \frac{Q_{\rm H}}{6.3 \times 10^{32}\, \rm [s^{-1}]} \left(\frac{T}{10^{4}\, \rm [K]}\right)^{0.45} \left(\frac{\nu}{\rm [GHz]}\right)^{-0.1}}.
    \label{freefree}
\end{equation}

\noindent Eq. (\ref{freefree}) is used in this paper to calculate the free-free radiation and is identical to that used in O17 (see Ep. 5 in that paper). Note that this equation is of the same form of the equation used to model free-free radiation in B02, but uses a different constant in the denominator of the production rate of Lyman continuum photons; we adopt the same value as O17, $6.3 \times 10^{32} s^{-1}$, whereas B02 used $\rm 5.495 \times 10^{32}s^{-1}$ (See Equation 1 in B02). B02 used their own simulation model of HII regions to calculate an average relation at 1.49~GHz to find $\rm 5.495 \times 10^{32} s^{-1}$. In this paper, we elect to use $\rm 6.3 \times 10^{32} s^{-1}$ since it comes from a purely theoretical understanding of free-free radiation. Like B02 and O17, we assume $T=10^{4}\,\rm  K$ which aligns with observations of HII regions (\citealt{anderson2009molecular}).

\subsubsection{Synchrotron Emission}

Synchrotron radiation is produced when electrons are accelerated to ultra-relativistic speeds. In the approach of B02, O17 and of this paper the dominant mechanism behind this acceleration is assumed to be CCSNe with a minor contributions from supernova remnants (SNR). Synchrotron radiation is calculated through Eq.~(\ref{sync_eqn}), which is identical to Eq.~(17) in B02:

\begin{multline}  \label{sync_eqn} 
    \frac{L_{\rm sync}(\nu)}{10^{23}\, [\rm W \,Hz^{-1}]} = 
    \biggl[ E_{\rm SNR} \left( \frac{\nu} {1.49\, [\rm GHz]} \right)^{-\alpha_{\rm SNR}} + \\ E_{\rm EI} \left( \frac{\nu}{1.49\,[\rm GHz]}\right)^{-\alpha_{\rm sync}}\biggr] 
    \frac{\nu_{\rm CCSN}}{[\rm yr^{-1}]},   
\end{multline}

\noindent where $E_{\rm SNR}$ is the energy contributed by SNRs, $E_{\rm EI}$ is the energy of electrons injected per SN event, $\alpha_{\rm SNR}$ and $\alpha_{\rm sync}$, are the radio luminosity power-law index for the frequency dependence of SNR and of electrons injected per SN event respectively, and $ \nu_{\rm CCSNe}$ is the rate of CCSNe.  

$\nu_{\rm CCSNe}$ is not assumed to be constant, but instead is calculated from the adopted IMF and the SFR of galaxies, 

\begin{equation}
    \frac{\nu_{\rm CCSNe}}{{[\rm yr^{-1}]}} =  \left(\frac{\alpha_{\rm CCSNe}}{[\rm M_{\odot}^{-1}]} \,\right)\,\left(  \frac{\rm SFR}{[\rm M_{\odot}\,yr^{-1}]} \right),
    \label{ccsn}
\end{equation}

\noindent 

\noindent where $\rm \alpha_{CCSNe}$ is the fraction of stars that undergo CCSNe per unit solar mass of stars formed. $\rm \nu_{CCSNe}$ is calculated for each galaxy's disk and bulge with a galaxy's total $\rm \nu_{CCSNe}$ being the sum of these two components.
It is a common assumption that the stars that eventually undergo CCSNe exist within the mass range of 8~$\rm M_{\odot} \lesssim$ M $\rm \lesssim$ 50~$\rm M_{\odot}$ \citep{heger2003massive,ando2003detectability,nomoto1984evolution,tsujimoto1997new}. Above this maximum mass, stars undergo hypernova, causing Gamma Ray Bursts (\citealt{van2007long}). Consequently $\alpha_{\rm CCSNe}$ can be expressed as follows; 
\begin{equation}
    \alpha_{\rm CCSNe} = \frac{\int_{8 \,\rm M_{\odot}}^{50 \,\rm M_{\odot}}\psi(M) dM}{\int_{0.1 \,\rm M_{\odot}}^{100\,\rm M_{\odot}}M\psi(M)\, {\rm d}M},
    \label{alpha_ccsn}
\end{equation}
\noindent where $\rm \psi(M)$ is the IMF. For a \cite{chabrier2003galactic} IMF this yields $\alpha_{\rm CCSNe}= 0.011 \rm \,M_{\odot}^{-1}$.

$\rm E^{SNR}$, $\rm E^{EI}$ and $\alpha_{\rm sync}$ are constants and we adopt the values $ E_{\rm SNR} = 0.0795 \times 10^{23}\rm \,W\,Hz^{-1}$, $E_{\rm EI} = 1.25 \times 10^{23}\rm \, W\,Hz^{-1}$ and $\alpha_{\rm sync} = 0.816$. These values are derived empirically and differ from those used in the B02 and O17 due to a different IMF being used in \textsc{Shark} compared to those works. The rest of this subsection is dedicated to the derivation of these constants.

B02 derive the total synchrotron emission in our galaxy using the result of \citet{berkhuijsen1984supernova}. They found the total synchrotron radiation observationally from our Galaxy at 408 MHz to be $L_{\rm MW,0.408GHz} = 6.1 \times 10^{21} \rm \,W\,Hz^{-1}$. Assuming a radio slope of $\rm \alpha = 0.8$ we convert this to the total synchrotron luminosity at 1.49 GHz: $L_{\rm MW,1.49GHz} = 2.13 \times 10^{21} \,\rm W \,Hz^{-1}$. It is possible to then find the average synchrotron luminosity per supernova event, $E^{\rm sync}$:

\begin{equation} 
E^{\rm sync} = \frac{L_{\rm MW,1.49\,GHz}}{\nu_{\rm CCSNe,MW}} = 1.24 \times 10^{23} \rm\, W\,Hz^{-1},
    \label{Esync}
\end{equation}

\noindent where $\nu_{\rm CCSN,MW}$ is the rate of CCSNe in the Milky Way. ${\nu_{\rm CCSNe,MW}}$ is assumed to be constant and we adopt ${\nu_{\rm CCSNe}} = 0.011\rm \, yr^{-1}$ which is calibrated with the Chabrier IMF used in \textsc{Shark} and uses the same  $\alpha_{\rm CCSNe}$ calculated above. B02 and O17 assume ${\nu_{\rm CCSNe}} = 0.015 \,\rm yr^{-1}$ which comes from \cite{cappellaro2001supernova} and uses a \citet{salpeter1955luminosity} IMF. 
It is this difference in ${\nu_{\rm CCSNe}}$ that results in the different constants used in this paper compared to those used in B02 and O17.

By assuming that the lifetime of synchrotron electrons is much smaller than the fading time of CCSNe rate (\citealt{volk1989correlation}) and that synchrotron radiation is the dominant loss mechanism, B02 shows that the synchrotron luminosity scales linearly with ${\nu_{\rm CCSNe}}$ with a constant $\rm E_{EI}$ for SFGs. This is also true for starbursts; to avoid losses from Inverse Compton scattering, the lifetime of electrons must be shorter in starbursts than in SFGs. On such a short timescale, $\nu_{\rm CCSN,MW}$ can assumed to be constant. B02 also shows that the synchrotron luminosity depends on the magnetic field, $B$, as $\propto B^{\alpha_{\rm sync}-1}$. We follow the same assumption of B02 that $\alpha_{\rm sync}$ is close to $1$ and thus the dependence on $B$ is weak and can be ignored. To quantify this statement, a change in $B$ of a factor of a $100$ (similar to the variations observed in high frequency peaker radio sources by \citealt{Orienti08} with $B$ in the range $\approx 0.01 - 0.1$~G) lead to a variation in $L_{\rm sync}$ of a factor $\approx 2$.

The B02 model also considers the contribution of SNRs, noting that other sources provide a negligible contribution. The average SNR synchrotron luminosity per SN event is:

\begin{equation}
 E_{\rm SNR} \simeq 0.06  E^{\rm sync}.
    \label{SNR}
\end{equation}

\noindent Eq. (\ref{SNR}) tells us that the contribution from SNR makes up about 6 percent of the synchrotron emission. The remaining 94 percent comes from electrons injected into the ISM and accelerated by magnetic fields. As previously derived, $E^{\rm sync} = 1.24 \times 10^{23}\rm \, W\,Hz^{-1}$ and so $E_{\rm EI} = 1.25 \times 10^{23}\,\rm W\,Hz^{-1}$, $E_{\rm SNR} = 0.0795 \times 10^{23}\,\rm W\,Hz^{-1}$.

SNRs have a spectrum which is modelled by $L_{\nu} = \nu^{\alpha_{\rm SNR}}$ where $\alpha_{\rm SNR} = 0.2 - 0.5$. The B02 model assumes that the radio slope of SNRs is constant at $\alpha_{\rm SNR} = 0.5$, which is less than the characteristic observed slope of the total non-thermal emission of SFGs ($\rm \alpha_{sync} = 0.8$). In order to compensate for this, the B02 model assumes that the spectrum for electrons injected into the ISM has a radio slope of $\alpha_{\rm EI} \simeq 0.9$ for an overall synchrotron radio slope of $\alpha_{\rm sync} = 0.8$. 

We differ in the approach to radio slopes. Like the B02 model we also assume that $\alpha_{\rm SNR} = 0.5$ but use a $\alpha_{\rm EI} = 0.816$ since this more accurately produces an overall slope of $\alpha_{\rm sync} = 0.8$. This slope gives an overall synchrotron slope of $\alpha_{\rm sync} = 0.8$ accounting for the flatter slope from the SNR contribution.

\subsection{Modelling the Radio Emission due to AGNs}\label{sec_agn_method}

In modelling the radio luminosities due to AGNs we start with the approach developed in F11 and extend that model to include the effects of synchrotron self-absorption. F11 models AGNs jet power and converts this to a bolometric luminosity and has been used successfully in \citet{griffin2019evolution}  and \citet{amarantidis2019first} to model AGN radio luminosities (the latter successfully modelling using multiple simulation suites including \textsc{Shark}). This model predicts the core jet power, but does not include extended emission (Such emission is only relevant in for very bright sources). We will now give a brief overview of the F11 model and its implementation into \textsc{Shark}, but we refer to the original paper for greater detail.

This model first calculates the power of the radio jets. To do so, the F11 model assumes that both BH spin and the accretion flow influence the power of the radio jet.  \citet{blandford1977electromagnetic} showed that the jet power, Q can be approximated by the BH mass ($M_{\rm BH}$), the magnitude of the BH spin ($\rm a$) and the strength of the poloidal magnetic field ($B_{\rm p}$):

 \begin{equation} 
  Q \propto B_{\rm p}^{2}\,M_{\rm BH}^{2}\,a^{2}.
     \label{eqn_bz_jet}
 \end{equation}

\noindent \textsc{Shark} v1.1 assumes that the BH spin parameter is constant at $\rm a =0.67$, as this corresponds to the standard radiation efficiency of $\rm 0.1$ \citep{bardeen1974rotating}. This assumption of constant BH spin is not realistic; it has been known to evolve with redshift, galaxy morphology and BH mass \citep{sesana2014linking, izquierdo2020galactic}. For that reason, the latest version of {\sc Shark} \citep{lagos2023} self-consistently tracks the development of BH spins, as they merge and accrete matter. We, however, elect to use \textsc{Shark} v1.1 as the FUV-to-FIR emission of galaxies has been investigated in detail \citep{lagos2019far,lagos2020physical}, and so instead assume a constant spin. We note that preliminary results of the same AGN radio emission model applied in the new version of {\sc Shark} produces qualitatively similar radio luminosity functions to what is found with a constant BH spin. However, we leave a full investigation of the results of the radio and FUV-to-FIR emission models in \textsc{Shark} v2.0 for future work.


$B_{\rm p}$ has a dependence on the azimuthal magnetic field strength, $B_{\phi}$ by $B_{\rm p} \approx (\rm H/R)\,B_{\phi}$. $\rm H/R$ is the ratio of BH accretion disk half-thickness to disk radius. This becomes important when considering whether the BH is in thin disk (TD), where $\rm H \ll R$, or advection-dominated accretion flow (ADAF) mode where $\rm H \sim R$ (\citealt{heckman2014coevolution}). Consequently the power of the jets (summed over both jets) in these two modes are (\citealt{meier2002grand}):

\begin{equation} 
  Q_{\rm ADAF} = 2 \times 10^{45} \, {\rm erg\,s^{-1}} \left(\dfrac{M_{\rm BH}}{10^{9}\,\rm M_{\odot}}\right) \left(\dfrac{\dot{m}}{0.01}\right)\,a^{2}, 
     \label{eqn_q_adaf}
\end{equation}
 
\begin{equation} 
  Q_{\rm TD} = 2 \times 10^{43} \, {\rm erg\,s^{-1}}\left(\dfrac{M_{\rm BH}}{10^{9}\,\rm M_{\odot}}\right)^{1.1} \left(\dfrac{\dot{m}}{0.01}\right)^{1.2}\,a^{2},
     \label{eqn_q_td}
\end{equation}

 \noindent where $\dot{m}$ is the dimensionless mass accretion rate $\dot{m} = \dot{M}/ \dot{M}_{\rm Edd}$. $\dot{M}$ is the accretion rate of the BH of each galaxy as simulated directly in \textsc{Shark} and $\dot{M}_{\rm Edd}$ is the Eddington Accretion rate, $\dot{M}_{\rm Edd} = L_{\rm Edd}/(0.1c^2)$.

F11 adopts \citet{heinz2003non} to compute the radio continuum luminosity at a specific frequency from the jet power. This model found that the flux of a jet scales as $M_{\rm BH}^{\xi_{1}}\dot{m}^{\xi_{2}}$ where $\rm \xi_{1} = \xi_{2} = 17/12$ for ADAF systems and $\rm \xi_{1} = 17/12, \xi _{2} = 0$ for radiation-pressure supported systems. Consequently it can be assumed for the luminosities of these two modes that

\begin{eqnarray}
L_{\rm ADAF} &\propto& (M_{\rm BH}\dot{m})^{1.42}
     \label{eqn_l_adaf_prop},\\
     L_{\rm TD} &\propto& M_{\rm BH}^{1.42}  \label{eqn_l_td_prop}.
\end{eqnarray}

 \noindent Thus the relations in Eqs.~\ref{eqn_l_adaf_prop} and \ref{eqn_l_td_prop} with Eqs.~\ref{eqn_q_adaf} and \ref{eqn_q_td}  can be combined such that the exponents of the former remain the same:

   \begin{eqnarray} 
L_{\rm ADAF}(\nu) &=& A_{\rm ADAF}\, Q_{\rm ADAF} \left(\dfrac{M_{\rm BH}}{10^{9}\,\rm M_{\odot}}\right)^{0.42} \left(\dfrac{\dot{m}}{0.01}\right)^{0.42}\nonumber \\
&& \cdot \left(\frac{\nu}{\rm 1.4 \, GHz}\right)^{\alpha_{\rm AGN}}, 
     \label{eqn_l_adaf}\\
L_{\rm TD}(\nu) &=& A_{\rm TD}\,Q_{\rm TD} \left(\dfrac{M_{\rm BH}}{10^{9}\,\rm M_{\odot}}\right)^{0.32} \left(\dfrac{\dot{m}}{0.01}\right)^{-1.2}\nonumber\\
&& \cdot \left(\frac{\nu}{\rm 1.4\,GHz}\right)^{\alpha_{\rm AGN}}, 
     \label{eqn_l_td}
 \end{eqnarray}

\noindent where $\rm A_{ADAF}$ and $\rm A_{TD}$ are normalisation coefficients. These are free parameters, and $\alpha_{\rm AGN}$ is the synchrotron power-law, which we set to $=-0.7$. F11 required that $\rm A_{TD} = 0.01 A_{ADAF}$, $\rm A_{ADAF} = 0.05$ for prolonged accretion and $\rm A_{ADAF} = 0.07$ for chaotic accretion. \citet{griffin2019evolution} compared three different sets of $\rm A_{ADAF}$, $\rm A_{TD}$ based on fitting to an AGN RLF at $\rm z = 0$ using the GALFORM model of galaxy formation. \citet{amarantidis2019first} take a similar approach with a $\rm \chi^2$ minimisation with AGN RLFs over multiple redshifts. For \textsc{Shark} this yielded $A_{\rm ADAF} = 1.3 \times 10^{-7}$ and $A_{\rm TD} = 8.0 \times 10^{-3}$. Notably, \citet{amarantidis2019first} found very different results for these normalisation coefficients across different galaxy formation models.

$A_{\rm ADAF}$ and $A_{\rm TD}$ remain free parameters within the radio emission model presented here and there is no correct method of finding them. In this paper we take a hybrid approach by performing a $\rm \chi^2$ minimisation procedure when compared with data of AGN RLF at $\rm z = 0$ at $\rm 1.4GHz$ from \citet{bonato2021new,ceraj2018vla,padovani2011microjansky,smolvcic2017vla} (see the left most panel of the middle row in Fig. \ref{fig_rad_lum_func_1d4}) to find $\rm A_{ADAF}$. We perform this fit is done without considering the effects of synchrotron self-absorption, which has negligible impact at $\rm 1.4GHz$. As \citet{amarantidis2019first} notes, due fewer BHs accreting in the radiative mode at lower redshifts, $\rm A_{TD}$ has little influence on the shape RLF at $\rm z = 0$. For this reason we assume that, like F11, $A_{\rm TD} = A_{\rm ADAF}/100$. We avoid taking the same procedure as \citet{amarantidis2019first} since, as we show in Section~\ref{subsec_RLF}, high redshift AGN RLFs are prone to contamination from galaxies dominated by SF. 

We adopt values of $A_{\rm ADAF} = 1.0 \times 10^{-5}$ and $A_{\rm TD} = 1.0 \times 10^{-7}$.

We extend the F11 model to include an empirical model of the synchrotron self-absorption which is expected to become increasingly important with decreasing frequency. We follow the empirical model of \citet{tingay2003investigation}, which characterised the synchrotron self-absorption of the nearby and very well-studied AGN, PKS 1718-649. To do so, they fitted a two component power-law relationship to the observed spectral data of PKS 1718-649, using a $\chi^2$ analysis.
We adopt this same two-component power-law and the parameters found in \citet{tingay2003investigation}:

  \begin{eqnarray}
L_{\rm AGN}(\nu)&=&\sum_{i = 1,2} L_{\rm F11}(\nu_i) \left(\dfrac{\nu}{\nu_i}\right)^{-(\beta_{i}-1)/2} \left[\dfrac{1-e^{-\tau_i,v}}{\tau_{i,v}} \right] ,\\
     \tau_{i,v} &=& \left(\dfrac{\nu}{\nu_{i}}\right)^{-(\beta_{i} +4)/2}, \label{eqn_ssa}
 \end{eqnarray}

\noindent where $L_{\rm F11}(\nu_i)=L_{\rm ADAF}(\nu_i)+L_{\rm TD}(\nu_i)$ is the luminosity from an AGN as calculated from the F11 model (using Eqs~\ref{eqn_l_adaf}~and~\ref{eqn_l_td}), $\rm \nu_i$ is the frequency at which the synchrotron optical depth is 1 and $\rm \beta_{i}$ is the power-law index, and $L_{\rm AGN}$ is the observed AGN luminosity after accounting for synchrotron self-absorption. We adopt the parameters from \citet{tingay2003investigation} that are $\rm v_{1} = 1.264$~GHz, $\rm v_{2} = 3.249$~GHz, $\rm \beta_{1} = 2.204$ and $\rm \beta_{2} = 1.905$.
\section{Results}\label{chp_results}

In this section we present the results of implementing the radio emission model into \textsc{Shark}. We start with the most general case of radio source number counts (Section \ref{sec_lightcone}) before increasing in specificity to radio luminosity functions (RLF) (Section \ref{subsec_RLF}), SFGs (Section \ref{sec_GAMA}) and (U)LIRGs (Section \ref{sec_ULIRGs}). Finally, in Section \ref{sec_qir} we present the results of investigations into the IRRC and $q_{\rm IR}$'s evolution with redshift and dependence on $\rm M_{*}$. These are all compared with observational results and show that the radio model in \textsc{Shark} is capable of reasonably reproducing these observational results.

\subsection{Galaxy number counts in radio frequencies}\label{sec_lightcone}

To compute number counts, we use the {\sc Shark} lightcone presented in Section~5 of \citep{lagos2019far}, which has an area of $107$~deg$^2$ and covers a redshift range $0\le z\le 6$. All galaxies with a dummy magnitude (calculated using a stellar mass-to-light ratio of $1$) $<30$~mag are included. We use the properties of the galaxies in the lightcone to compute the radio continuum emission following Sections~\ref{sec_rad_method}~and~\ref{sec_agn_method}. Using Eqn. 4 from \citet{driver2010quantifying} with the area and redshift range this lightcone has an approximate cosmic variance of $1.11$\%. However, if we only focus on the low-redshift part of the lightcone ($0\le z\le 0.25$), cosmic variance is expected to be $\approx 8$\%.

\begin{figure*}
    \centering
    \includegraphics[trim=4mm 1mm 2mm 3mm, clip,width=0.99\textwidth]{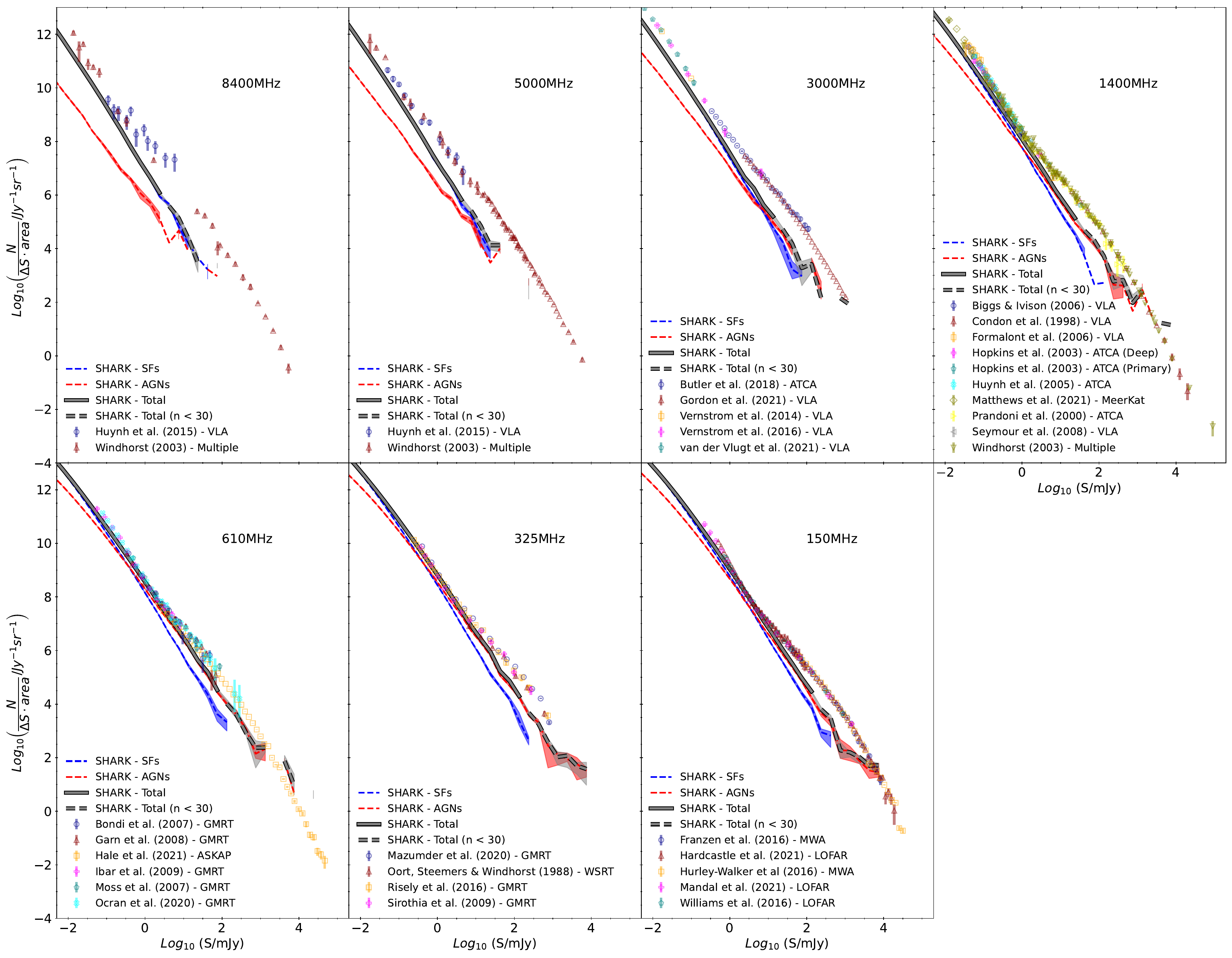}
    \caption{Number counts of galaxies at the labelled frequency. Emission from the \textsc{Shark} SF model (ie the B02 model) is shown in blue and emission from the \textsc{Shark} AGN model (ie the F11 model) is shown in red. The combined total of these two emissions is shown in the thick, grey line, solid grey line shows where the total number of sources is $\rm \geq 30$ and the dashed grey line shows where $\rm < 30$. The shaded regions show the $\rm 1-\sigma$ error found using bootstrapping.  Data points show counts from different observations from a variety of sources as compiled in \citet{tompkins2023cosmic}. This includes data from \citet{huynh2015atlas, windhorst2003microjansky} (8400~MHz and 5000~MHz) \citet{butler2018xxl, gordon2021quick, vernstrom2014deep, vernstrom2016deep, van2021ultradeep} (3000~MHz), \citet{biggs2006catalogue, condon1998nrao, hopkins2003phoenix, huynh2005radio, matthews2021source, prandoni2000atesp, seymour2008star,windhorst2003microjansky} (1400~MHz), \citet{bondi2007vvds,garn2008610,hale2021rapid,ibar2009deep,moss2007610,ocran2020deep} (610~MHz), \citet{mazumder2020characterizing, oort19886, riseley2016deep, sirothia2009325} (325~MHz) and \citet{franzen2016154, hardcastle2021contribution, hurley2017galactic, mandal2021extremely, williams2016lofar} (150~MHz).}
    \label{fig_n_v_z_land}
\end{figure*}

Fig.~\ref{fig_n_v_z_land} shows the radio source counts for the \textsc{Shark} model compared with observations for frequencies from  $\rm 150\, MHz$ to $\rm 8400\, MHz$. The emission solely from SF and AGNs are shown in the dashed lines (blue and red respectively) and their combined total radio emission is shown in the thick grey line. The solid grey line shows where \textsc{Shark} has more than 30 galaxies in the sample and the dashed grey line is where there are fewer than 30 galaxies in the total \textsc{Shark} sample, so is less statistically robust. Errors calculated using 
bootstrapping are also shown in the shaded regions.
Generally, radio source counts can be characterised by the combination of two curves; one dominated by SF and one dominated by AGNs. The intersection of these two curves occurs at brighter fluxes with higher frequencies; SF dominates at $\rm log_{10}(S/mJy) \lesssim 0.75$ at 8400~MHz whereas at 150~MHz this is a order of magnitude fainter at $\rm log_{10}(S/mJy) \lesssim -0.75$. This is not unexpected, as higher frequencies have a smaller synchrotron emission contribution (due to the frequency dependence of the synchrotron emission), requiring ever brighter AGNs to dominate over SF as the frequency increases. 


Number counts are compared with observational results compiled by  \citet{tompkins2023cosmic} which includes a comprehensive compendium of radio source data from a variety of papers, frequencies and instruments. They also found this two humped distribution where AGNs dominate at the bright end and SF at the faint end.
Overall there is a reasonable level of agreement between \textsc{Shark} and the observations. However, there are some tensions and trends that are worth noting. For the purpose of this analysis we define radio faint fluxes as $-2 \leq \rm log_{10} (S/mJy) \leq 0$ and radio bright fluxes as $0 \leq \rm log_{10} (S/mJy) \leq 2$. We find the difference in between the number counts of the observations and \textsc{Shark} in dex. Here, the relevant number counts for \textsc{Shark} correspond to the total line (summing over the contribution from both SF and AGNs). Note that this difference is calculated for each flux bin, but below we quantify the differences between {\sc Shark} and observations by computing the median difference in the flux range we define above for faint and bright radio sources. 

For faint radio sources, \textsc{Shark} predicts a number of galaxies that is within $1$~dex of what is observed: 
this difference is greatest at $\rm 8400\,MHz$ of $-0.70$~dex, but this decreases towards lower frequencies with   
$-0.03$~dex at $\rm 325\,MHz$,  and $-0.23$~dex at 150~MHz. We consider {\sc Shark} to be in reasonable agreement with observations of the number counts of faint radio sources. We therefore conclude that the model of radio emission from SF introduced in Section ~\ref{sec_rad_method} can adequately reproduce the radio source counts in \textsc{Shark}.

A greater source of tension is seen in the radio bright regime. Again across all frequencies \textsc{Shark} predicts fewer objects at fixed flux than observations. This tension is greatest at $8400$~MHz with a difference of $-1.53$~dex and at $5000$~MHz a difference of $-1.06$~dex, but improves at lower frequencies with $-0.28$~dex at $\rm 610\,MHz$, $-0.26$~dex at $\rm 325\,MHz$, and $-0.22$~dex at $\rm 150\,MHz$.

\begin{figure*}
    \centering
    \includegraphics[trim=1mm 3mm 0mm 2mm, clip,width=0.99\textwidth]{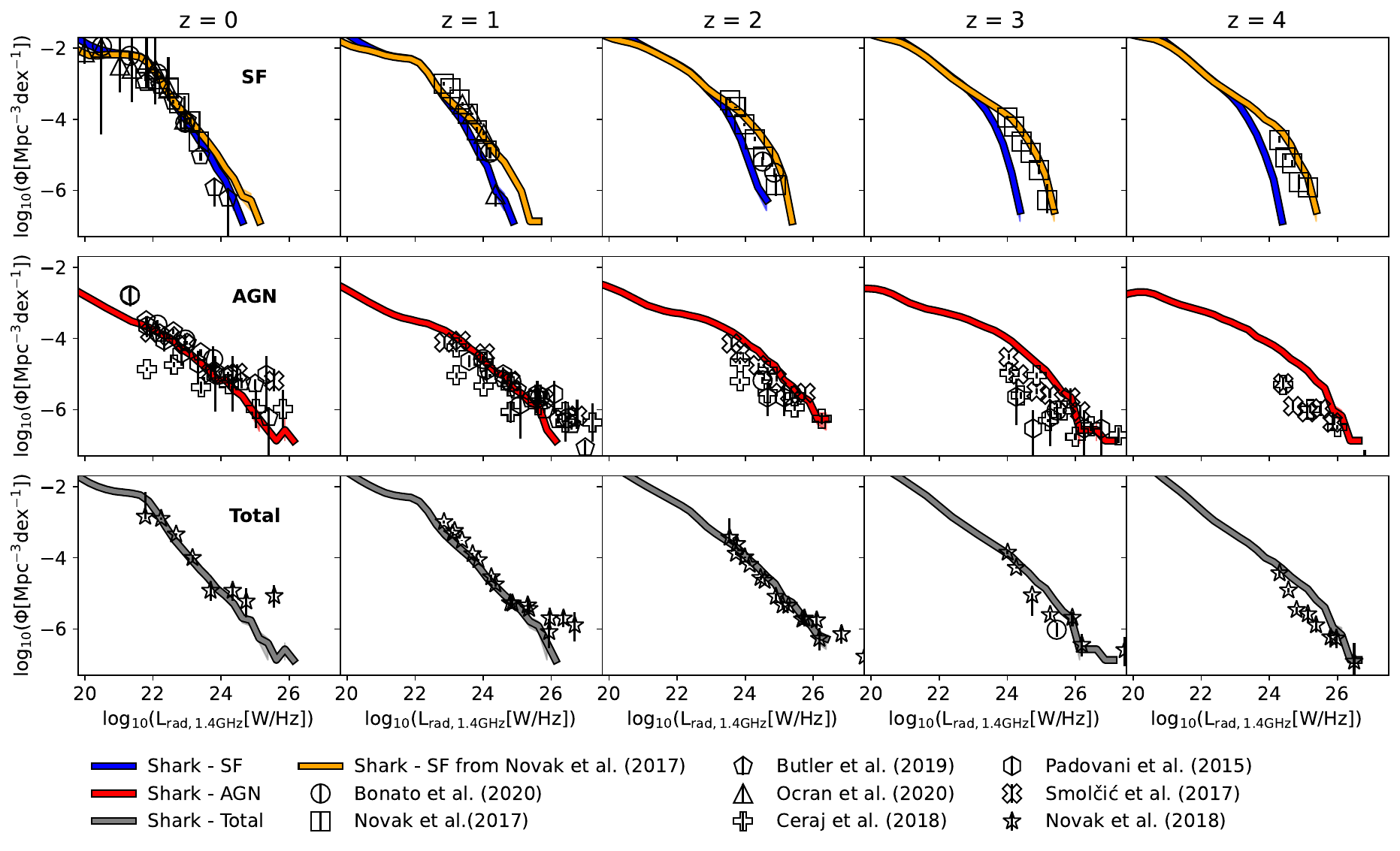}
    \caption{The RLF at 1.4~GHz associated to SF (top panels), to AGNs (middle panels) and the total (bottom panels). Each column is a different redshift from $z = 0$ to $4$, as labelled. The shaded regions show the $\rm 1\,\sigma$ error found from bootstrapping. Comparisons are made with observational data of galaxies classified as SFGs in the top panels, AGNs in the middle panels, and the total RLF in the bottom panels; \citet{bonato2021new}, \citet{butler2019xxl}, \citet{novak2017vla} and \citet{ocran2019cosmic} (for SFGs), \citet{padovani2011microjansky, ceraj2018vla,smolvcic2017vla,ceraj2017cosmic} for AGNs and \citet{novak2018constraints} for the total. In the top panels, the blue lines show the true luminosities associated to SF in {\sc Shark}, while the orange line shows SFGs as identified using the classification from \citet{novak2017vla}. The difference between the two lines can be attributed to AGN contamination in the latter classification.}
    \label{fig_rad_lum_func_1d4}
\end{figure*}

As AGNs dominate the radio bright regime, the tension 
arises primarily from the model of AGN emission. To understand why the model falls below the observations at the bright end, we studied the redshift distribution of the bright sources in {\sc Shark}. We focus on the $1.4$~GHz sources as those are the ones we can compare with the RLFs at different redshifts with a range of observations. We find that most of the galaxies with a flux $>100$~mJy in {\sc Shark} in the lightcone are at $z<0.5$. These are galaxies that have AGN luminosities $ > 10^{25.5}$~W/Hz. This tension is also apparent in the radio bright regime of the AGN RLF in Fig~\ref{fig_rad_lum_func_1d4}. We discuss this further in Section \ref{subsec_RLF}.


Two aspects are important to understand why our model leads to an underprediction of the number counts of bright sources in the radio regime. The first one is that the constants $A_{\rm ADAF}$ and $A_{\rm TD}$ used in Eqs.~\ref{eqn_l_adaf}~and~\ref{eqn_l_td} are calibrated to reproduce  intermediate luminosity AGNs in the $\rm 1400\,MHz$ RLF at $z = 0$, which is where the constraints are strongest. The second aspect is that in this work extended radio emission is not modelled, which is expected to dominate the flux in galaxies of $1.4$~GHz luminosities $>10^{25}$~W/Hz (e.g. \citealt{gendre2013relation}). Hence, it is not surprising that the number counts fall below the observations at the very bright end. This is something we will revisit once we introduce a model for the extended radio emission of AGNs.

\subsection{Radio luminosity functions across cosmic time} \label{subsec_RLF}

We now examine \textsc{Shark}'s performance in reproducing the radio luminosity functions, which is a traditional benchmark to which previous emission models have been compared (\citealt{lagos2019far,somerville2012galaxy,wilman2008semi,bonaldi2019tiered,murphy2012star}). 
We compare \textsc{Shark} results to observations at $\rm 1.4\,GHz$ (Section~\ref{sec_rlf_1d4}) and $\rm 150\,MHz$ (Section~\ref{sec_rlf_150}) and across cosmic time. Note here that these results do not employ the lightcone introduced in Section ~\ref{sec_lightcone}; instead using \textsc{Shark}'s native simulated box.

\subsubsection{The 1.4~GHz luminosity function} \label{sec_rlf_1d4}

Fig.~\ref{fig_rad_lum_func_1d4} shows the RLF at 1.4GHz associated to SF in galaxies (top panels), AGNs (middle panels) and the total (bottom panels) across different redshifts. 

At $\rm z = 0$, \textsc{Shark} compares favourably with the observed data of SFGs and the total radio. By construction \textsc{Shark} will agree well with observations of the AGN RLF at $\rm z = 0$ as we used those observations to fit for $\rm A_{ADAF}$. Unlike the AGN RLF at z = 0, the agreement with SFGs and in the total RLF was not guaranteed. This  
gives us confidence the SF emission model produces reasonable results.

Note that in the AGNs and total RLF, {\sc Shark} produces too few galaxies brighter than 
$\rm log_{10}(L_{\rm rad,1.4GHz}/\rm W \,Hz^{-1}) \gtrsim 25$. This is related to the tension seen in the predicted number counts in the radio bright regime in Fig. ~\ref{fig_n_v_z_land}. This tension indicates that the AGN luminosity model is under-predicting the number density of very bright AGNs at low redshifts. We attribute this to the lack of extended AGN emission in the model, which dominates AGN emission at $\rm log_{10}(L_{\rm rad,1.4GHz}) \gtrsim 25\,\rm  W Hz^{-1}$ \citep{gendre2013relation}. These luminosities are also close to the limit of the simulated box. As there are very few SF galaxies for $\rm log_{10}(L_{\rm rad,1.4GHz}) \gtrsim 25\,\rm  W Hz^{-1}$, the lack of radio bright AGN emission naturally means that the total RLF under-predicts as well at these luminosities.


This good agreement between \textsc{Shark} and the observations continues into the second column for the SF RLF. For AGNs and total this is also true for ${\rm log}_{10}(L_{\rm rad,1.4GHz}) \lesssim 26\,\rm  W Hz^{-1}$.

For the total RLF, there is good agreement with observations at $\rm z = 2$ and $z=3$. At $z=4$, however, there is an overabundance of galaxies with $24.5 \lesssim \rm log_{10}(L_{\rm rad,1.4GHz}/W Hz^{-1})\lesssim 25.5$ in {\sc Shark} compared to observations, though outside that luminosity range {\sc Shark} agrees well with observations. 
The reasonable agreement we get, specially for the total RLF at $1.4\,\rm GHz$ shows that the model performs well even though it has been calibrated at $z=0$ only.

The level of agreement we see in the top and middle panels, however, is not as good as that seen in the bottom panels. 
At $\rm z \ge  2$ we start to see a persistent under prediction in the RLF associated with SF in galaxies (top panels), and we see the tension with observations increasing from $z=2$ to $z=4$. Note that most of the observations are limited to the range
$\rm 25 \lesssim log_{10}(L_{rad,1.4GHz}/ W\,Hz^{-1})\lesssim 27$.
Interestingly, over this same luminosity range, the \textsc{Shark} AGN RLF (middle panels) predicts an $\emph{over}$ abundance of galaxies compared with observations. 
At $\rm log_{10}(L_{rad,1.4GHz}/ W\,Hz^{-1})\gtrsim 27$, \textsc{Shark}'s predictions for the AGN RLF are in good agreement with observations.
The tension seen here is something that is not unique to \textsc{Shark}. \citet{jose2024understanding} modelled the synchrotron emission of SF galaxies in the semi-analytic model \textsc{GALFORM} and found good agreement with observations for $\rm z < 2$, but that the model produced synchrotron luminosities much lower than observations at $\rm z \gtrsim 2$.
Below we discuss the tension described here for the RLF of SFGs and AGNs and propose that can be explained in our model by a large fraction of the radio-quiet AGN being mis-classified as SFGs in the observations.

There are numerous caveats to the observations particularly at high redshifts. This includes the underestimation of errors associated with the observations. For each observation we have included the error provided by the papers, however many of these only include Poisson error. This does not account for other sources of error like cosmic variance. Results from \citet{driver2010quantifying} estimate the cosmic variance of the area covered by some of these surveys to be as high as $\rm 50\%$.
Another caveat is the determination of redshift observationally. These are mostly done using photometry and there is a high probability for catastrophic failure; for example \citet{novak2018constraints} estimates $\rm \sim 12.5\%$ chance of catastrophic failures at $\rm z \geq 1.5$. The percentage of catastrophic failures is expected to increase with increasing redshift.

Caveats aside, we considered multiple solutions to addressing the tension between the RLFs of SF and AGNs at these high redshifts. 
One possible option for the tension seen in the top panels for the contribution of SF to the total RLFs could be that the SFRs are too low in {\sc Shark}. This is not the case and \citet{lagos2023} in fact showed in their supplementary material (their Fig.~6) that both {\sc Shark} v1.1 and v2.0  reproduce the observed SFR function from $z=0$ to $z=8$ very well.
Another option is that the IMF may be varying, as suggested in other SAMs \citep{baugh2005can,lacey2008galaxy,lacey2016unified}.
We invoked a top-heavy IMF in {\sc Shark} by post-processing the galaxies to assume different CCSNe rates for SF associated with galaxy disks and starbursts. For disks we continue to assume a Chabrier IMF, but for starbursts we use the top-heavier IMF adopted in \citet{lacey2016unified}, which gives us $\rm \alpha_{CCSNe} = 0.022M_{\odot}^{-1}$ in Eq.~(\ref{ccsn}).
Because starbursts are more prevalent at higher redshifts in {\sc Shark} (\citealt{lagos2018shark,lagos2020physical}), the higher $\rm \alpha_{CCSNe}$ is expected to increase $L_{\rm rad,1.4GHz}$ with increasing redshift. This exercise indeed helps to improve the agreement with observations. However, a full test of this solution would require a varying IMF to be included self-consistently within {\sc Shark}.
In addition, 
\citet{lagos2019far} found that the UV-to-IR LFs at different cosmic times could be reproduced well by {\sc Shark} assuming a universal Chabrier IMF, which makes us reluctant to use a varying IMF to reproduce the RLFs.

We also considered cosmic variance as a source of error that could explain this tension. To investigate this we calculated the $\rm 1.4 GHz$ RLF but for smaller volumes of the simulated box (specifically volumes of $\rm 5^{3} Mpc^{3}$). While this did increase the error associated with each RLF, it did not bridge the gap between \textsc{Shark} and the observations. We conclude that cosmic variance on its own does not explain the tension in the RLF at $\rm z \sim 3,4$.

\begin{figure*}
    \centering
    \includegraphics[trim=1mm 4mm 0mm 2mm, clip,width=0.99\textwidth]{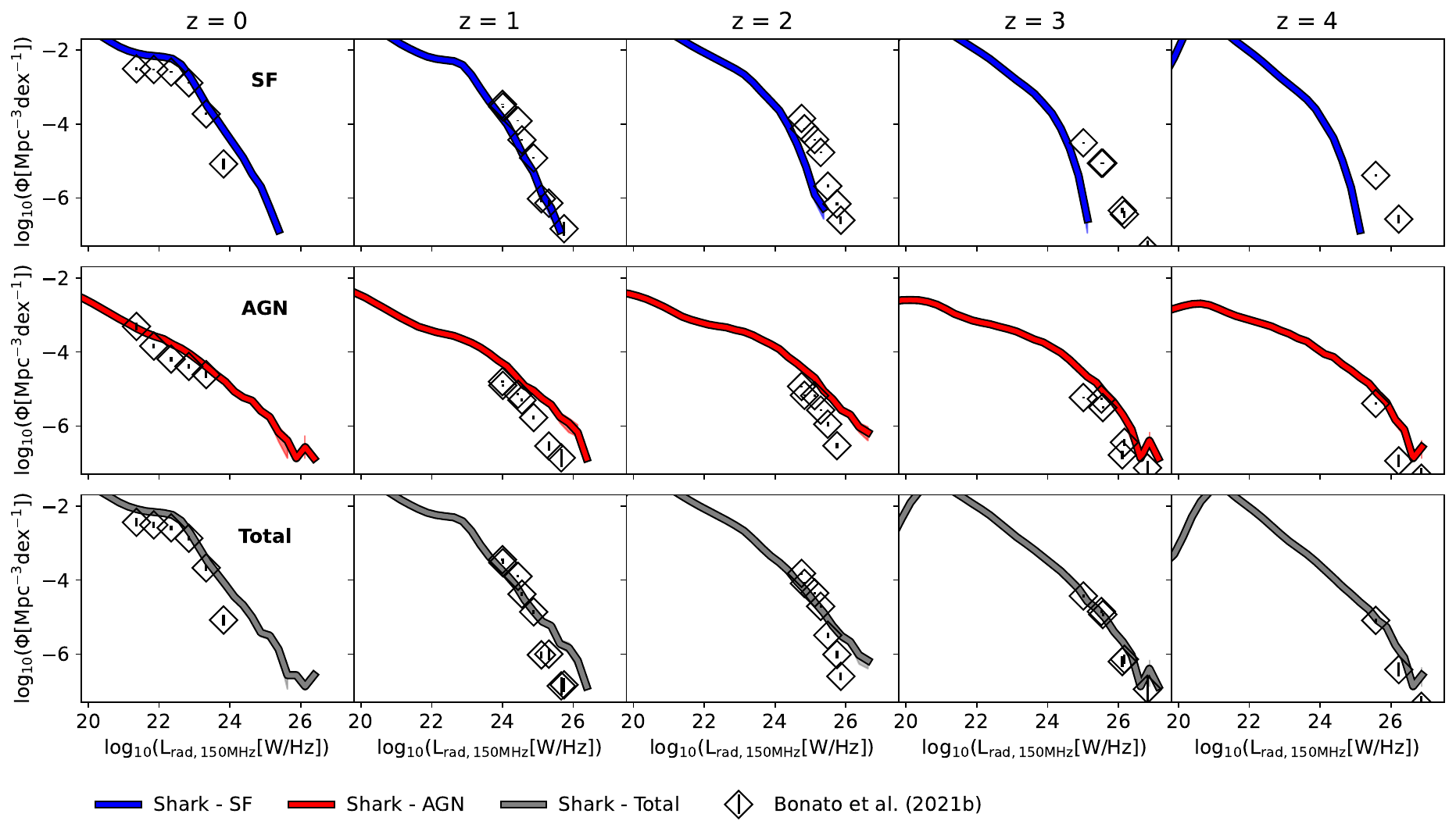}
    \caption{As in Fig.~\ref{fig_rad_lum_func_1d4} but for $150\,\rm MHz$. Observational results are from \citet{bonato2021lofar}. We remark that the errorbars associated to the observational datapoints correspond to Poisson uncertainties only.}
    \label{fig_rad_lum_func_150}
\end{figure*}

We propose instead that this tension can be resolved if we consider the way galaxies are selected to be SF or AGNs in observations.
Observational studies do not have the luxury of being able to easily distinguish between light purely from AGNs and SF in any one galaxy. Instead observational studies classify galaxies as being either SF or AGN, often in a binary way. 
%
This becomes increasingly more difficult to do at higher redshifts as less information (in both the sense of observed radio morphology and the ability to data match with other surveys at different frequencies) is available for these galaxies. \textsc{Shark} is in the comparatively privileged position of not having such limitations, and can also apply the same selections employed in observations to classify galaxies.

\citet{novak2017vla} classify and remove radio loud AGNs from their SF RLF using a single criterion first used in \citet{delvecchio2017vla} based on a source's $L_{\rm rad,1.4GHz}$ and SFR to classify it as an AGN:

\begin{equation} \label{eqn_novak}
{\rm log}_{10} \left(\dfrac{L_{\rm rad,1.4GHz}}{\rm SFR_{IR}}\right) > 22 \times (1+z)^{0.003}
\end{equation}

\noindent where $L_{\rm rad,1.4GHz}$ is in units of $\rm W\, Hz^{-1}$ and $\rm SFR_{IR}$ in $\rm M_{\odot}\, yr^{-1}$. We thus take the total RLF in {\sc Shark}, and remove every {\sc Shark} galaxy that complies with Eq.~(\ref{eqn_novak}), irrespective of their individual AGN or SF activity ($L_{\rm rad,1.4GHz}$ used in Eq. \ref{eqn_novak} is a galaxy's total radio luminosity with contributions from both SF and AGNs).  The remaining would be equivalent to what \citet{novak2017vla} report as ``SFGs'', and thus would be a fair comparison with their reported RLFs.
Note that $\rm SFR_{IR}$ is determined using the SFR-IR relation from \citet{kennicutt1998global}. For completeness we derive a galaxy's $\rm SFR_{IR}$ in the same way in {\sc Shark}, but note that a similar result is seen if we instead use the intrinsic galaxy's SFR.

The resulting sample's RLF is shown by the orange lines in the top panels of Fig.~\ref{fig_rad_lum_func_1d4}. The agreement with the observations is now excellent. 
This shows that \textsc{Shark} can successfully reproduce observational results when following the same selection method as such observational results, and that it is crucial to attempt to apply the same selection methods as employed in observations to understand how successful {\sc Shark} is at reproducing observational results.

We speculate that the classification of galaxies into SF/AGN dominated bins would also be responsible for the over abundance in {\sc Shark} galaxies in the AGN RLF. In high redshift surveys, the sub-mJy population consists of mainly SF-dominated sources, but with some AGN contribution. Empirical AGN diagnostics would classify such sources as solely SF ignoring their composite nature.

\subsubsection{The 150~MHz luminosity function}\label{sec_rlf_150}

Fig.~\ref{fig_rad_lum_func_150} shows the 150~MHz RLF. We compare with the observational results of \citet{bonato2021lofar}.

Focusing first on the top left panel of Fig.~\ref{fig_rad_lum_func_150}, we see that {\sc Shark} reproduces well the $\rm z = 0$ RLF associated with SFGs. We remind the reader that in the top panels of Fig.~\ref{fig_rad_lum_func_150} we only include the emission associated with SF for the {\sc Shark} predictions. The predicted normalisation by {\sc Shark} is a factor of $3$ too high compared with the observed number density at the brightest bin at $\rm log_{10} (L_{rad,150MHz}) \sim 24 W Hz^{-1}$. However, these observations only account for Poisson errorbars and hence are likely very under-estimated, specially at the bright-end (see discussion in Section \ref{sec_rlf_1d4} about cosmic variance). At higher redshift (top second to fifth panels in Fig.~\ref{fig_rad_lum_func_150}), we see similar trends as those discussed for the SF $\rm 1.4GHz$ RLF; \textsc{Shark} predicts an increasingly lower number density than observations of bright galaxies with 
increasing redshift. Note that observations are only able to constrain the bright-end of the $150\,\rm MHz$ RLF at $z\ge 1$; deeper observations at $150\,\rm MHz$ are needed to measure where the break of the RLF is and the slope below the break. As with $\rm 1.4GHz$, the tension at the bright-end is likely due to contamination from radio-quiet AGNs. \citet{bonato2021lofar} use data from \citet{best2023lofar}, which classifies galaxies as AGN-dominated using four different SED fitting methods. The SED modelling applied in {\sc Shark} only considers the contribution from stars and dust attenuation/re-emission in the FUV-to-FIR, hence we cannot apply the same selection criteria as done in \citet{best2023lofar} and test for AGN contamination. However, the similarity of the trends here and those presented in Fig.~\ref{fig_rad_lum_func_1d4} makes us suspect that AGN contamination in the SFGs selection may be an issue here too.

For the AGN RLF (middle panels in Fig.~\ref{fig_rad_lum_func_150}) we obtain reasonable agreement between {\sc Shark} and the observations across different redshifts. The best agreement is at $\rm z = 0$, however with increasing redshift we see an opposite trend to that seen in the SF RLF (albeit to a smaller magnitude); \textsc{Shark} predicts a higher number density of bright AGNs at higher luminosities than what is seen in the observational data. This can be seen at $\rm z = 1$, where \textsc{Shark} over-predicts compared to the entire observational set, but more evidently at $\rm log_{10} (L_{rad,150MHz}) \gtrsim 25 W Hz^{-1}$.
We speculate that this is likely due to the inverse effects of AGN contamination seen in the SF RLF; that the removal of AGNs from the observational data due to their misclassification as SF dominated has lead to the observational data under representing the whole set. Note that we do not see the same tension at radio bright luminosities  at $\rm z = 0$ for AGNs that is seen in Figs. ~\ref{fig_n_v_z_land} and ~\ref{fig_rad_lum_func_1d4}. There is simply no observational data at $\rm 150MHz$ that probes these high luminosities at this redshift.

The level of agreement between the total RLF and observations is overall better (shown in the bottom panels of Fig.~\ref{fig_rad_lum_func_150}), except at the very bright end of each redshift. The better agreement naturally results from the SF contribution being a bit too low and the AGNs one being a bit too high, compensating to some degree. This lends support to our interpretation of the tension in the top and middle panels of Fig.~\ref{fig_rad_lum_func_150} being due to SF/AGN separation difficulties in the observations.

We note that preliminary results obtained using  \textsc{Shark} v2.0 \citep{lagos2023} show  qualitatively similar radio luminosity functions as seen in Figs. ~\ref{fig_rad_lum_func_1d4} and ~\ref{fig_rad_lum_func_150}, however a full analysis of these results are left for future work.

\begin{figure*}
    \centering
    \includegraphics[trim=7mm 7mm 6mm 6mm,
    clip,width=0.99\textwidth]{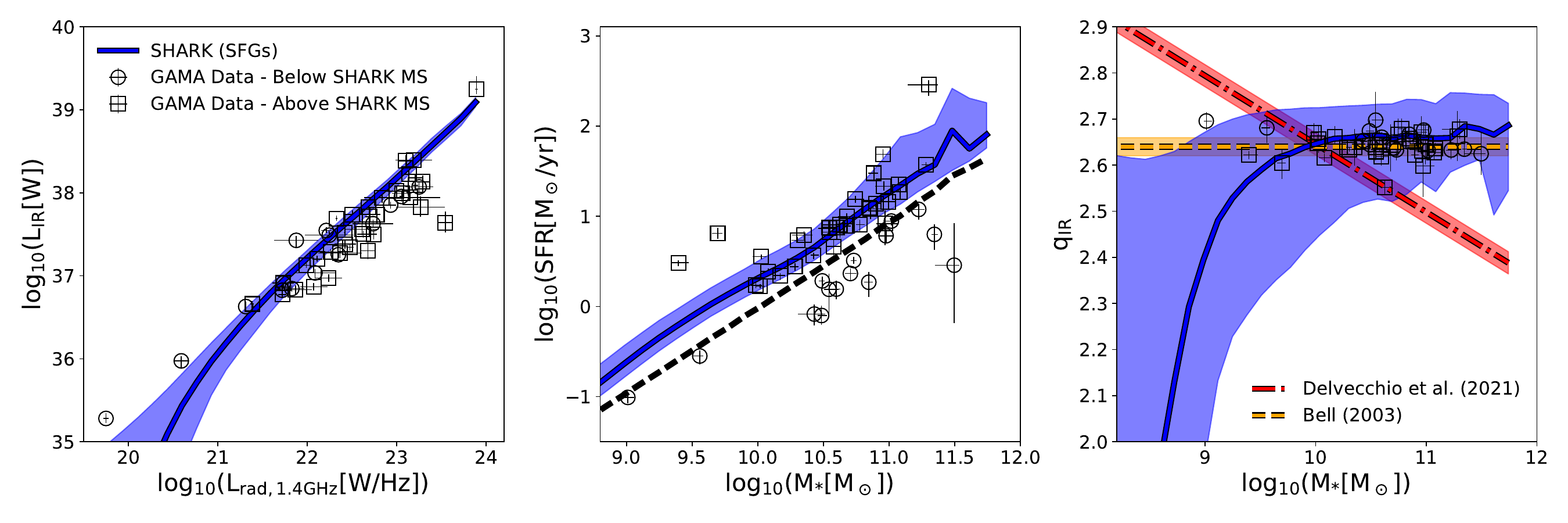}
    \caption{{\it Left panel:} The IR-1.4~GHz luminosity relation for SFGs in {\sc Shark} at $z=0$. The medians and the $16^{\rm th}-84^{\rm th}$ percentile ranges are shown as a line and shaded region, respectively. Symbols show individual GAMA galaxies from the combined GAMA catalogue of \citep{davies2017galaxy} (1.4~GHz luminosities come from this catalogue) and \citet{bellstedt2020} (IR luminosities come from this catalogue). Circles are GAMA galaxies that occur above the \textsc{Shark} MS and squares those that occur below.
    {\it Middle panel:} the SFR-stellar mass plane of SFGs in {\sc Shark}. Solid line with the shaded region and symbols are as in the left panel. 
    All galaxies above the black dashed line are considered SFGs in {\sc Shark} (see text for details). {\it Right panel:} 
    The $q_{\rm IR}$-stellar mass plane. Solid line with the shaded region and symbols are as in the left panel. 
    The red dot-dashed line shows the $q_{\rm IR}$-stellar mass relationship found in \citet{delvecchio2021infrared} (Fig.~14 in that paper), while the orange dashed line shows the constant $q_{\rm IR}$ found by \citet{bell2003estimating}.  
    } 
    \label{fig_GAMA_model_plts}
\end{figure*}


\subsection{Radio continuum scaling relations in the local Universe}\label{sec_GAMA}

We now turn to \textsc{Shark}'s performance with respect to SFGs in the local Universe. To test this we compare with observational results from the Galaxy And Mass Assembly (GAMA) survey (\citealt{Driver2022galaxy}). We select only SFGs as they are thought to be the main population driving the observed IRRC. Note that in this case we {\it only} include the radio continuum associated to SF and ignore the AGN contribution. To define our population of SFGs within \textsc{Shark} we first define the main sequence (MS) using a linear fit in the space $\rm log_{10}(SFR)-log_{10}(M_{*})$ for central galaxies with stellar masses between $\rm 9 \leq log_{10}(M_{*}/M_{\odot}) \leq 10$. These mass limits are chosen to avoid resolution limitations on the lower end, and AGN quenching at the higher end. We define SFGs then as those that have $\rm log_{10}(SFR)$ that is $>-0.3$~dex from the main sequence. The latter is inspired by the main sequence having a scatter of $\rm 0.2 - 0.3$~dex at $z\approx 0$ in observations (\citealt{Davies19,popesso2023main}) and in simulations \citep{katsianis2019, Davies19}. As in Section ~\ref{subsec_RLF}, the \textsc{Shark} sample  is not sourced from the lightcone, rather \textsc{Shark}'s native simulation box $\rm z = 0$ is used.

Fig.~\ref{fig_GAMA_model_plts} shows three scaling relations connecting the radio continuum emission at $1.4$~GHz with other galaxy properties for SFGs in {\sc Shark} and observations.
The observed data here comes from the GAMA analysis of the $1.4$~GHz luminosity of galaxies presented in \citet{davies2017galaxy}, which were sourced from the Faint Images of the Radio Sky at Twenty cm (FIRST) survey (\citealt{becker1995first}). The IR luminosity comes from the SED fitting analysis of the same galaxies presented in \citet{bellstedt2020}. The latter is defined as the total luminosity that is re-radiated in the IR, which is the same definition used in {\sc Shark}, and hence are comparable. The data here is for galaxies at $z \le 0.06$ and thus we compare it with {\sc Shark} predictions at $z=0$. 
We show GAMA galaxies that occur above the \textsc{Shark} MS as circles and those that are below as squares (although galaxies below MS are only slightly below).

The left panel of Fig.~\ref{fig_GAMA_model_plts} shows $\rm L_{IR}$  as a function of $L_{\rm rad,1.4GHz}$ (i.e. the IRRC).
Both populations agree well with each other despite the measurements being determined in different ways. The scatter seen in the GAMA data is larger than the intrinsic scatter in {\sc Shark}, but that is not surprising as we have not include any potential errors in obtaining luminosities, and the range of SFRs of GAMA galaxies is larger than that of SFGs in {\sc Shark} (see middle panel of Fig.~\ref{fig_GAMA_model_plts}). The agreement shown here was not guaranteed as the IR and radio luminosities in {\sc Shark} are modelled independently as described in Section~\ref{cpt_method}. 

The left panel of Fig. \ref{fig_GAMA_model_plts} shows that in {\sc Shark} the median mostly follows a straight line between $\rm log_{10}(L_{\rm IR})$ and $\rm log_{10}(L_{\rm rad,1.4GHz})$ - however, in 
low luminosity galaxies, $L_{\rm rad,1.4GHz} \lesssim 10^{21}\,\rm  W/Hz$, the trend changes so that there is less $L_{\rm IR}$ per unit $L_{\rm rad,1.4GHz}$ than in brighter galaxies. There are only two GAMA galaxies in that regime, which mostly follow the extrapolation of the trend seen in brighter galaxies. Because of the small number of observations it is hard to establish whether there is tension or not with {\sc Shark} in the faint regime. We see no discernible difference in the distribution of GAMA galaxies that are above \textsc{Shark}'s MS and those that are below in this panel.

The middle panel of Fig.~\ref{fig_GAMA_model_plts} shows the SFR-$\rm M_{*}$ plane. This plot shows that the GAMA sample used here (which only includes those galaxies with FIRST detections) are good representations of SFGs in {\sc Shark}. 28.9\% of GAMA galaxies fall below the dashed line, which marks our threshold to classify galaxies in {\sc Shark} as being SFGs. However, these galaxies are only mildly below our MS definition in {\sc Shark}, and the left and right panels of Fig.~\ref{fig_GAMA_model_plts} shows that effectively they follow the same relations as the galaxies above MS.

The right panel of Fig.~\ref{fig_GAMA_model_plts} shows $q_{\rm IR}$ against stellar mass for SFGs. This panel allows for a relative comparison of different $q_{\rm IR}$ results. GAMA, \citet{bell2003estimating} and \textsc{Shark} all agree in the stellar mass range $\rm 10^{10} - 10^{12}\,\rm M_{\odot}$. This is also the regime 
%
where the majority of the data from GAMA and \citet{bell2003estimating} comes from.

Fig.~\ref{fig_GAMA_model_plts} also shows the $q_{\rm IR}$-stellar mass relation reported by \citet{delvecchio2021infrared} from a set of observations spanning a wide range of stellar masses and redshifts. The authors concluded that there was very little redshift dependence of the $q_{\rm IR}$-stellar mass relation, so we include it in this figure.
There is apparent 
tension between the data from \textsc{Shark} and local Universe observations with the relation reported in \citet{delvecchio2021infrared}. We do not find the same dependence on stellar mass that \citet{delvecchio2021infrared} found. We investigate this tension thoroughly in Section \ref{sec_qir}. 

There are very few galaxies from GAMA or \citet{bell2003estimating} with stellar masses $\rm < 10^{10} \,\rm M_{\odot}$ so meaningful comparisons with observations cannot be made over this mass range. Below a stellar mass of $10^{10} \,\rm M_\odot$, {\sc Shark} predicts a sharp decrease of $q_{\rm IR}$ with a rapid increase in the scatter of the relation. Some galaxies remain in $q_{\rm IR}\approx 2.6$ as seen from the $84^{\rm th}$ percentile, but the majority have $q_{\rm IR}<2.5$ at a stellar mass $<10^{9} \,\rm M_\odot$. 
This is driven by the modelling of the IR and radio continuum from SF.
In \textsc{Shark}, the vast majority of these galaxies have their optical depth dominated by birth clouds rather than the diffuse ISM. The optical depth of BCs depends linearly on the ISM metallicity (see Eq.~(6) in \citealt{lagos2019far}). Overall this means as $\rm Z_{gas}$ drops with $\rm M_{*}$ so  does the optical depth resulting in less UV emission being absorbed and re-emitted in the IR. This means that there is less IR luminosity per unit SFR in these low-mass galaxies in {\sc Shark}. The scatter around the median is driven by some low-mass galaxies still having significant diffuse ISM attenuation (those that are close to $q_{\rm IR}\approx 2.6$) and those with insignificant diffuse ISM attenuation (those with $q_{\rm IR} \ll 2.6$). 

In \textsc{Shark}, the attenuation due to the ISM is sampled from the attenuation curves reported by \citet{trayford2019fade}. The high end of this sampling can result in galaxies having a non-negligible ISM optical depth, up to $\rm \tau_{ISM} \sim 0.15$ at these stellar masses (see Fig.~3 in \citealt{lagos2019far}) even though their ISM metallicity can be slightly sub-solar, effectively cancelling out the decrease in optical depth in the BCs.

This trend in IR not balanced by a similar decrease in radio emission leading to an overall decrease in $q_{\rm IR}$. The model of radio emission due to SF in \textsc{Shark} is predicated on such radio emission being a perfect tracer of the rate of CCSNe. However, \citet{bell2003estimating} argued that the linearity of the IRRC is proof that radio emission is not a perfect tracer of radio emission in low-luminosity galaxies. Further \citet{chi1990implications} suggested that escape of cosmic ray electrons in lower mass galaxies can explain the linearity of the IRRC. Such a downturn in $q_{\rm IR}$ as seen in Fig.~\ref{fig_GAMA_model_plts} for {\sc Shark} is due to the lack of modelling of this escape.

There is an overall lack of comprehensive observational data at the stellar masses to which this effect becomes significant. Hence, at this stage we leave this as a caveat of our modelling in {\sc Shark} and stress that the model seems to apply well to galaxies with stellar masses $\gtrsim 10^{9.7}\,\rm M_{\odot}$ and leave it for future work to test the model in the low-mass regime. We note that the latter will be possible in the near future thanks to surveys such as Evolutionary Map of the Universe 
(EMU; \citealt{norris2021evolutionary}) being carried out with ASKAP, and the VLASS (\citealt{Lacy2020Karl}).

\subsection{Radio continuum scaling relations at high redshift} \label{sec_ULIRGs}

We now compare the capabilities of \textsc{Shark} with observations of galaxies at $1\lesssim z\lesssim 2.5$. We do this by compared with galaxies classified as being (U)LIRGs. Specifically we compare with the sample presented in \citet{lo2015combining}. As in the previous section, we only consider the radio continuum associated with SF here.

In \citet{lo2015combining} LIRGs are classified as having a flux at $\rm 24 \mu\rm m $ ($\rm S_{24 \mu m}$) $\sim 0.2 - 0.5$ mJy at $z = 0.76 - 1.05$. ULIRGs are classified as having $S_{\rm 24 \mu m} \sim  0.14 - 0.55$ mJy at $z = 1.75 - 2.4$. This differs from the usual definition of LIRGs having IR luminosities in the range  $\rm 10^{11} - 10^{12} L_{\odot}$ and ULIRGs in the range $\rm 10^{12} - 10^{13} L_{\odot}$. We adopt the same definition of \citet{lo2015combining} for a fair comparison. For this we use the lightcone presented in \citet{lagos2019far} in their Section~5 (the same used in Section~\ref{sec_lightcone}), after applying the radio continuum model presented in this paper. 

\begin{figure*}
    \centering
    \includegraphics[trim=7mm 7mm 5mm 4mm, clip,width=0.99\textwidth]{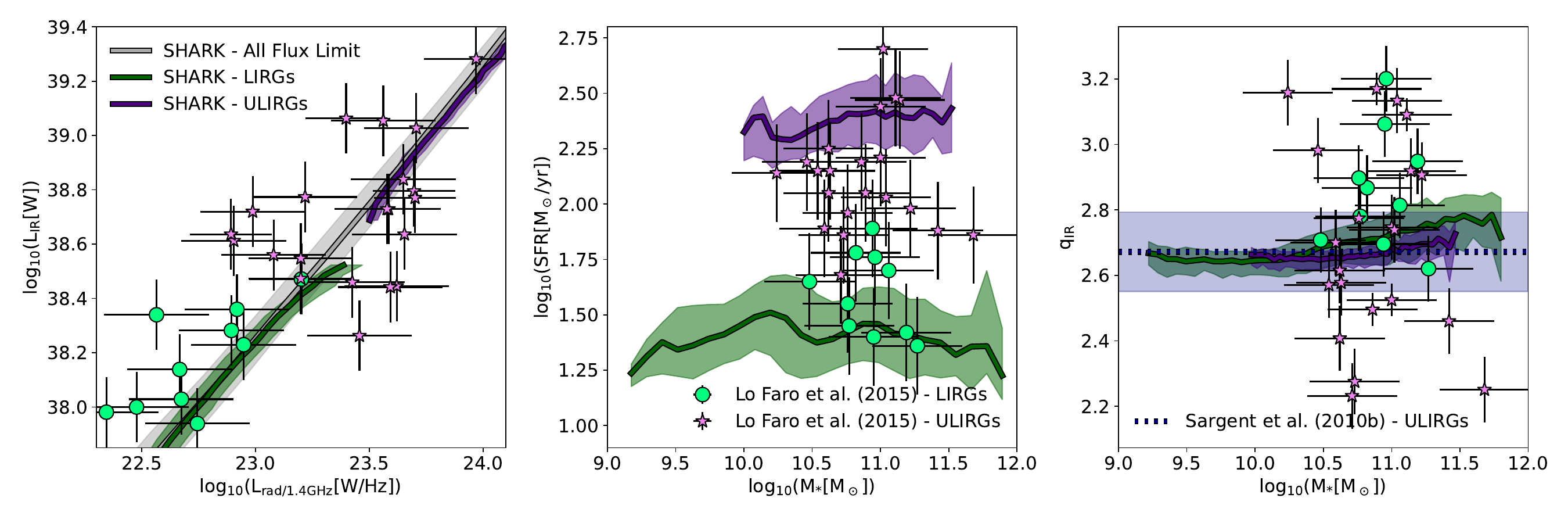}
    \caption{As in Fig.~\ref{fig_GAMA_model_plts} but for LIRGs (green) and ULIRGs (purple) galaxies from {\sc Shark} (lines with shaded regions) and \citet{lo2015combining} (symbols). The right panel also includes the median $q_{\rm IR}$ found for ULIRGs in \citet{sargent2010no}. For {\sc Shark}, we classify galaxies ad LIRGs and ULIRGs following the same classification adopted in \citet{lo2015combining}: all galaxies with a $24\,\rm mu m$ flux in the range $0.15-0.45$~mJy are selected and those in the redshift ranges $0.76-1.05$ and $1.75-2.4$ are considered LIRGs and ULIRGs, respectively. In the left panel we also show the median and $16^{\rm th}-84^{\rm th}$ percentile ranges for all galaxies in {\sc Shark} with a $24\,\rm mu m$ flux in the range $0.15-0.45$~mJy regardless of their redshift (grey line and shaded region, respectively).}  
    \label{fig_lo_faro_plts}
\end{figure*}

Fig.~\ref{fig_lo_faro_plts} shows the IR-radio continuum, SFR-stellar mass and $q_{\rm IR}$-stellar mass relations for LIRGs and ULIRGs. 
In all panels, LIRGs are shown in green and ULIRGS in purple with the lines showing the median relations and the shaded region the $\rm 1-\sigma$ percentile range of \textsc{Shark} galaxies. Observations of individual galaxies from \citet{lo2015combining} are shown with symbols coloured in the same way. The grey line in the left panel of Fig.~\ref{fig_lo_faro_plts} shows the population of galaxies selected with only the flux limit ($S_{24 \mu m} \sim  0.14 - 0.55$ mJy) from \textsc{Shark} and no redshift selection.

Compared with the observational results from \citet{lo2015combining}, \textsc{Shark} broadly agrees. \citet{lo2015combining} showed that a clear IRRC exists for both LIRGs and ULIRGs and \textsc{Shark} agrees with this result. 


In this same panel, for LIRGs in \textsc{Shark}, there appears to be a flattening at the brighter end of the slope ($ L_{\rm rad,1.4GHz} \sim 10^{23.25}\,\rm W\, Hz^{-1}$). A similar phenomena appears to be occurring at the faint end for ULIRGs at $L_{\rm rad,1.4GHz} \sim 10^{23.5}\, \rm W\, Hz^{-1}$ where the slope appears steeper than at higher luminosities. These differences in slope are due to the redshift limits imposed on these populations. This is exemplified by the grey line showing all galaxies in \textsc{Shark} within the flux limit. This line bridges the gap between LIRGs and ULIRGs. Hence, this gap and differences in slopes are driven by the specific selection we are applying here to identify LIRGS and ULIRGS.

The middle panel of Fig.~\ref{fig_lo_faro_plts} shows the SFR-stellar mass relation for LIRGs and ULIRGs. The stellar mass of the observed galaxies are determined using SED fits using two different codes: \citet{fadda2010ultra} use \textsc{HYPERZ} (\citet{bolzonella2000photometric} while \citet{lofaro2013complex} and \citet{lo2015combining} use \textsc{GRASIL} (\citealt{silva1998modeling}).
The stellar masses shown here were found using \textsc{HYPERZ} (with the difference between the two methods being insignificant).

Fig.~\ref{fig_lo_faro_plts} shows that the stellar mass ranges of LIRGs and ULIRGs in \citet{lo2015combining} are in broad agreement with those found in \textsc{Shark} despite no stellar mass selections in either dataset.
Similarly to stellar mass, \citet{lo2015combining} use two methods of determining SFR. The first is SED fitting using \textsc{GRASIL} 
and the other uses the SFR-$\rm L_{IR}$ relation from \citet{kennicutt1998global}. We show here the latter estimate.  
Again, there is broad agreement between \textsc{Shark} and \citet{lo2015combining} on the SFRs of both galaxy populations. 
Although it does appear like the {\sc Shark} distribution is more bimodal, that stems from the results shown being medians and percentile ranges, rather than individual data points as we show for 
\citet{lo2015combining}.
The median \textsc{Shark} result is within the margin of error for most of the observations. However is that {\sc Shark} reproduces well the range of SFRs and stellar masses seen in LIRGs and ULIRGs, and the fact that ULIRGs have on average higher SFRs than LIRGs at fixed stellar mass.


The SFR of ULIRGs in \textsc{Shark} is also nearly half an order of magnitude higher than that observed in \citet{lo2015combining}. This discrepancy is not necessarily concerning as the sample 
from \citet{lo2015combining} is very small (10 LIRGs and 21 ULIRGs). 
\citet{lagos2020physical} compared the SFRs and stellar masses of sub-mm bright galaxies in {\sc Shark} with a much larger a complete sample of observed galaxies (of many hundreds), finding that \textsc{Shark} was able to reproduce the observations well within the uncertainties. 
Therefore it is likely that the discrepancy here is driven by the difficulty in comparing with such a small sample.

The right panel of Fig.~\ref{fig_lo_faro_plts} shows the dependence of $q_{\rm IR}$ on stellar mass for LIRGs and ULIRGs. In addition to the \citet{lo2015combining} data, we also show the median $q_{\rm IR}$ of \citet{sargent2010no} for ULIRGs as a dotted line.  \citet{sargent2010no} studied 1,692 ULIRGs and 3004 ``IR bright'' sources out to $\rm z = 2$ from  the VLA-COSMOS "Joint" Catalogue, and found $q_{\rm IR} = 2.672 \pm 0.121$ that is independent of redshift. They found no evolution of $q_{\rm IR}$ with redshift but did not test for a possible dependence on $\rm M_{*}$. 
\textsc{Shark} finds no appreciable dependence of $q_{\rm IR}$ on $\rm M_{*}$ for (U)LIRGs and the median $q_{\rm IR}$ is within the margin of error of that found by \citet{sargent2010no}. 

In {\sc Shark}, LIRGs have a higher median $q_{\rm IR}$ than ULIRGs at the same stellar mass. On the surface, this seems counterintuitive since by definition, ULIRGs have an increased $\rm L_{IR}$ compared with LIRGs. However, as shown in Fig. \ref{fig_lo_faro_plts}, ULIRGs have an appreciably higher SFR. This SFR leads to an increased $\rm L_{rad,1.4GHz}$ which balances the increased $\rm L_{IR}$ in ULIRGs and results in a similar $q_{\rm IR}$.

In {\sc Shark}, the ULIRG selection applied here leads to galaxies whose SFR primarily comes from the starburst star formation mode, while LIRGs have a much higher contribution from the disk star formation mode. {\sc Shark} assumes both SFR modes follow the same relation between the surface density of SFR and molecular gas, except for the normalisation of the burst mode being $10$ times higher than that of the disk mode. The motivation for this was discussed in \citet{lagos2018shark} but shortly is inspired by sub-millimeter galaxies having a higher SFR efficiency per unit molecular gas mass than normal star-forming galaxies in observations. Thus, the different modes of star formation can lead to galaxies having different SFR but the same amount of molecular and dust mass, impacting the $L_{\rm IR}-\rm SFR$ relation. This difference in contribution to SFR from starbursts is what leads to a slightly higher $q_{\rm IR}$ (by $0.1$ dex at most) at fixed stellar mass for LIRGs compared to ULIRGs in {\sc Shark}.

LIRGs with stellar masses $\lesssim 10^{10.5}\,\rm M_{\odot}$ have a higher contribution to their SFR arising from the starburst mode compared with more massive LIRGs. This results in $q_{\rm IR}$ of those lower mass LIRGs being similar to the $q_{\rm IR}$ of ULIRGs in {\sc Shark}. This transition of the SF mode that dominates the total SFR leads to the weak dependence of $q_{\rm IR}$ on stellar mass for LIRGs that is seen in {\sc Shark}.

The difference in $q_{\rm IR}$ between LIRGs/ULIRGs was seen in the results of \citet{lo2015combining}; LIRGs were found to have a higher median $q_{\rm IR}$ than ULIRGs, however a small sample size in that paper meant it was unable to robustly conclude this.

Note that to see the difference between $q_{\rm IR}$ and stellar mass for LIRGs and ULIRGs in {\sc Shark} would require very large sample sizes for each population in the observations, and precise measurements of $L_{\rm IR}$ and $L_{\rm rad,1.4GHz}$ as to see an average difference of $\approx 0.1$ in $q_{\rm IR}$. This is  unattainable with current observations, but the upcoming SKA combined with multi-wavelength observations that allow SED fitting and robust measurements of the IR luminosity, are likely to be sufficient to test the predictions made here.

We also investigated using the usual definition of LIRGs having IR luminosities in the range  $\rm 10^{11} - 10^{12} L_{\odot}$ and ULIRGs in the range $\rm 10^{12} - 10^{13} L_{\odot}$ in {\sc Shark}, and  found no important  
differences from the results presented here.

\begin{figure*} 
    \centering
    \includegraphics[trim=2mm 3mm 2mm 2mm, clip,width=0.99\textwidth]{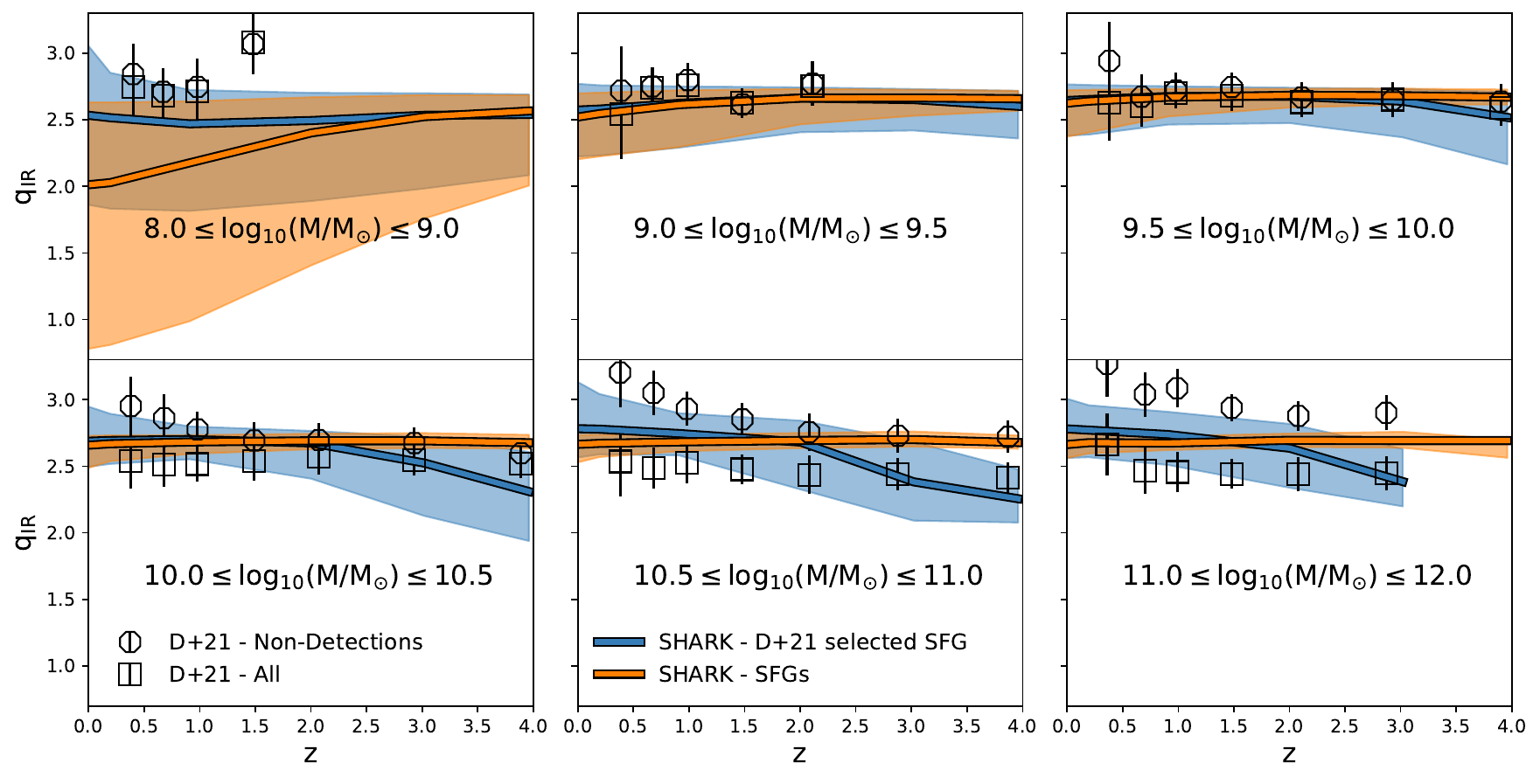}
    \caption{$q_{\rm IR}$ for as a function of redshift for different stellar mass ranges, as labelled. In orange, we show SFGs in {\sc Shark} as defined in Section~\ref{sec_GAMA}, and considering the radio continuum associated with SF only. In blue, we show the resulting $q_{\rm IR}$ when considering the total radio continuum (coming from both SF and AGNs) and after applying the same galaxy selection as in \citet{delvecchio2021infrared}, whose observations are shown as symbols. 
    Solid lines show the median $q_{\rm IR}$ and the shaded area shows the $\rm 1-\sigma$ percentile range. We show two sets of observations from \citet{delvecchio2021infrared}, median stacked values of non-detections as octagons and the weighted average $q_{\rm IR}$ of detections and non-detections as squares. These two sets should bracket the plausible range of $q_{\rm IR}$ values in the observational sample.}
    \label{fig_delv}
\end{figure*}
\subsection{The IR-radio correlation and its dependence on redshift and stellar mass} \label{sec_qir}

We now turn to the IRRC and what \textsc{Shark} can tell us about its evolution with redshift and the influence, if any, that stellar mass has on it. Fig.~\ref{fig_delv} shows $q_{\rm IR}$ as a function of redshift for different stellar mass ranges for two different ways of selecting \textsc{Shark} SFGs. Orange shows SFGs in \textsc{Shark} (as defined by their distance to the main sequence as described in Section~\ref{sec_GAMA}), and only including radio continuum emission associated with SF; specifically we set each galaxy's luminosity to that produced solely by the B02 model with no consideration of AGNs.
We compare that with the results in blue, which is a population of galaxies in {\sc Shark} whose radio emission is dominated by SF following the methodology of \citet{delvecchio2021infrared}. In this case, we use the total radio continuum emission of galaxies, which includes both SF and AGNs. As with  the results of  Sections ~\ref{subsec_RLF} and ~\ref{sec_GAMA}, here, we do not employ the lightcone. We compare with the observational results from \citet{delvecchio2021infrared}, which are presented as symbols. The two sets of symbols show 
the stacked values of non-detections as octagons and the weighted average $q_{\rm IR}$ of detections and non-detections as squares.

The way \citet{delvecchio2021infrared} identified galaxies dominated by SF in the radio is as follows. 
A total $q_{\rm IR}$ is calculated from a galaxy's total radio emission.
Then, the main dependence of $q_{\rm IR}$ on redshift is removed by fitting the function $\rm q_{IR} \propto (1+z)^{\alpha}$ and the subtracting the fitted function to the individual $q_{\rm IR}$ values. In effect we are calculating the distance between $q_{\rm IR}$ of individual galaxies and the fitted $\rm q_{IR} \propto (1+z)^{\alpha}$ function.
In \citet{delvecchio2021infrared} this is only done on the two most massive bins (ie $\rm 10.5 \leq log_{10}(M/M_{\odot}) \leq 11$ and $\rm 11 \leq log_{10}(M/M_{\odot}) \leq 12$) as these are the mass bins that are most complete. Though we are not limited by this completeness problem, we do the same here to allow for a fair comparison. 
Having removed the main trend with redshift, a histogram is created of $q_{\rm IR}$ in the two most massive bins (see Fig.~10 in \citealt{delvecchio2021infrared}). We then identify the peak or mode of this distribution, $q_{\rm peak}$. It is assumed that this peak and all galaxies with $q_{\rm IR}$ greater than it are dominated by SF and that the distribution is symmetric about $\rm q_{peak}$. In \textsc{Shark} $\rm q_{peak} = 2.85$ for both $\rm 10.5 \leq log_{10} (M/M_{\odot}) \leq 11$ and $\rm 11 \leq log_{10} (M/M_{\odot}) \leq 12$. We can then take the distribution of galaxies with $q_{\rm IR}>q_{\rm peak}$ and mirror it about $q_{\rm peak}$ and fit a Gaussian to the resulting distribution. From this Gaussian we find $q_{\rm thres}  = q_{\rm peak} - 2\sigma$ where $\rm \sigma$ is the standard deviation of the fitted Gaussian. In \textsc{Shark} $\rm sigma = 0.4$ for  $\rm 10.5 \leq log_{10} (M/M_{\odot}) \leq 11$ and $\rm \sigma = 0.35$ for $\rm 11 \leq log_{10} (M/M_{\odot}) \leq 12$. This is a larger dispersion than that found in \citet{delvecchio2021infrared} of $0.2$ and $0.23$ respectively. $q_{\rm thres}$ defines the dividing line between SF and AGNs; galaxies with $\rm q_{IR}>q_{\rm thresh}$ are classified as being SFGs and those below are AGN. The method is then repeated, removing more galaxies with $q_{\rm IR} < q_{\rm peak}$ until the median $q_{\rm IR}$ is unchanged within the uncertainties. In \citet{delvecchio2021infrared} only two iterations are performed before this condition is reached as such we only perform this process twice on \textsc{Shark} galaxies for similar results.
The methodology for galaxies with stellar masses $< 10^{10.5}\,\rm M_{\odot}$ is the same with  one key difference. The $q_{\rm IR}$-redshift function above is extrapolated to lower stellar masses rather than re-fitted.

Having removed AGNs using a similar methodology as \citet{delvecchio2021infrared}, we create the median and $1 - \sigma$ percentile range of those {\sc Shark} galaxies (shown with blue lines and shaded regions in Fig.~\ref{fig_delv}). This is the result that is comparable with the observations reported in \citet{delvecchio2021infrared}.
In this comparison we find excellent agreement with {\sc Shark} in all but a few observational points within the margin of error of the \citet{delvecchio2021infrared} results. This is quite remarkable given the complexity of both the radio continuum emission model, and the many steps involved in the selection of SFGs. 
One caveat to this comparison is that the sample used in \citet{delvecchio2021infrared} is restricted to 'blue' star forming galaxies defined from their optical colours. In the results presented here we make no such optical identification, but note that we found similar results when such optical identification was included.

Some of the conclusions we can draw from {\sc Shark} are different though than those drawn by \citet{delvecchio2021infrared}. By putting all the trends shown by the blue lines in Fig.~\ref{fig_delv}, we conclude that the redshift dependence of $q_{\rm IR}$ on redshift is stronger than that on stellar mass in {\sc Shark}, which is the opposite to what \citet{delvecchio2021infrared} concluded. The data, however, does agree with {\sc Shark} within the uncertaintites pointing out that these conclusions are subject to potentially important systematic effects.

The most striking aspect of Fig.~\ref{fig_delv} is the comparison between the two \textsc{Shark} populations presented.  \textsc{Shark} predicts a clear redshift evolution for galaxies with stellar masses in the range $10^8-10^9\,\rm M_{\odot}$, which is not recovered by the \citet{delvecchio2021infrared} selection applied to {\sc Shark} galaxies. This is a consequence of the low $q_{\rm IR}$ galaxies being removed as suspects of AGN contamination. In this case, however, this population of low-mass low $q_{\rm IR}$ galaxies is purely driven by SF, and a result of the modelling included in {\sc Shark} as described in Section~\ref{sec_GAMA}. 
At higher stellar masses ($>10^9\,\rm M_{\odot}$)  {\sc Shark} the intrinsic $q_{\rm IR}$ (outlined by the orange lines in Fig.~\ref{fig_delv}) show little to no redshift evolution and little stellar mass dependence. In other words, $q_{\rm IR}$ is close to constant (with some scatter) for all galaxies with stellar masses $\gtrsim 10^9\,\rm M_{\odot}$. This differs from the picture one could draw from studying the recovered $q_{\rm IR}$ evolution once we apply the method of selecting SFGs in the observations (blue lines in Fig.~\ref{fig_delv}). 

The key difference here is the influence of AGNs within the \citet{delvecchio2021infrared}-selected SFGs. We therefore conclude that the evolution of $q_{\rm IR}$ with z as seen in some observations (e.g \citealt{ivison2010blast,ivison2010far,magnelli2015far,delhaize2017vla}) may largely be driven by AGN contamination from radio-quiet AGNs. We call this ``radio-quiet'' AGN contamination, because the galaxies that have very bright radio AGN (typically refer to as ``radio-loud'' AGNs) are the ones that are confidently removed from the galaxy sample once we follow the selection method of \citet{delvecchio2021infrared}.

The conclusion above is very important as it tells us that using $q_{\rm IR}$ in some form to select SFGs, to then measure a $q_{\rm IR}$ may lead to biased results that do not completely remove contaminants. Using independent methods to remove AGNs (ie \citet{cook2024devils}) is likely a better choice, though this becomes more difficult to do as we move to samples at high redshift which have less multi-wavelength information from which to independently tag AGNs. This also shows that including all sources of radio continuum emission in {\sc Shark} is key, even if to evaluate galaxy samples that are dominated by SF given how difficult it is to completely remove the AGN contribution. Finally, the analysis here also shows the importance of following the same selection criteria employed in observations (or at least as closely as possible) to truly assess how well the model performs against observations.

\section{Conclusion} \label{sec_conclusion}

We have introduced a model of radio continuum emission associated to SF and AGNs in the semi-analytic model of galaxy formation \textsc{Shark}. We build off the results from \citet{lagos2019far}, which successfully modelled the UV-FIR emission using \textsc{Shark} to find the IR emission of galaxies.

We use the approach developed in \citet{bressan2002far} to model emission from SF, which includes synchrotron and free-free emission. Synchrotron is modelled as proportional to the rate of CCSNe with a minor contribution from SNe remnants. Free-free emission is modelled as proportional to the production rate of ionising photons (as a proxy for the number of free electrons). 

To model radio emission from AGNs we adopt the \citet{fanidakis2011grand} model. This model finds the radio luminosity of AGNs as a function of the power of radio jets. This model depends on the BH mass, accretion rate and spin. The latest version of {\sc Shark} \citep{lagos2023} predicts all these properties for each BH in the simulation, however, for the version of {\sc Shark} we use here (that of \citealt{lagos2018shark}), only the BH mass and accretion rate are predicted for individual BHs. For the spin we adopt $0.67$, which is equivalent to assuming a standard radiation efficiency of $0.1$.

Below we summarise our main findings and conclusions:
\begin{itemize}
    \item We show that this model is capable of reproducing a variety of key observations: (i) radio source counts over seven different frequencies from $8.4$~GHz to $150$~MHz (Fig.~\ref{fig_n_v_z_land}) (to better than $10\%$ in most bands and frequencies and to better than $30$\% overall); (ii) the total RLFs at $1.4$~GHz and $150$~MHz from $z=0$ to $z=4$ (bottom panels in Figs.~\ref{fig_rad_lum_func_1d4} and \ref{fig_rad_lum_func_150}); (iii) the scaling relations between the IR luminosity-$1.4$~GHz luminosity and stellar mass of SFGs in the local universe (Fig.~\ref{fig_GAMA_model_plts}) and at high-redshift for LIRGs and ULIRGs (Fig.~\ref{fig_lo_faro_plts}). 
    \item We find that the exact methodology of separating SFGs and AGNs has an impact in the level of agreement we obtain with observations. In particular, we show in Fig~\ref{fig_rad_lum_func_1d4} that applying the same method that is described in \citet{novak2017vla} to separate SFGs and AGNs leads to a good agreement with the RLFs reported there for the two galaxy populations. However, we note that the intrinsic prediction of the contributions to the RLF from SF and AGNs being different to what ends up being associated to either population after applying the method  in \citet{novak2017vla}. This shows that a direct comparison between the intrinsic prediction and the observations would have led us to conclude that poor agreement existed between the two populations.
    \item We see a similar tension between the $150$~MHz RLFs of SFGs and AGNs reported in observations and the intrinsic prediction in {\sc Shark} for the contribution to the total RLFs from SF and AGNs. Because the method adopted by \citet{bonato2021lofar} to separate SFGs and AGNs at $150$~MHz is more difficult to reproduce (as we lack a model for the mid-IR emission of AGNs in {\sc Shark}), we do not attempt to do that in this paper, and leave it for future work to assess. However, the results obtained at $1.4$~GHz indicate that AGN contamination needs to be treated carefully when comparing with observations.
    \item We investigate the relationship between $q_{\rm IR}$, redshift and stellar mass. We first do this by computing $q_{\rm IR}$ for \textsc{Shark} galaxies considering only the radio continuum emission associated with SF. The latter results in no evolution of $q_{\rm IR}$ with redshift and no dependence on $M_{\rm*}$ for galaxies with $M_{\rm*} \gtrsim 10^{9}\,\rm  M_{\odot}$. However, if we include the radio continuum emission of AGNs and follow the method of \citet{delvecchio2021infrared} to remove AGN galaxies, we find that the resulting sample has significant AGN contamination leading to an apparent evolution of $q_{\rm IR}$ with redshift. The resulting $q_{\rm IR}$ agree well with the observations of reported in \citet{delvecchio2021infrared}, but deviate from the intrinsic predictions in {\sc Shark} for radio continuum associated with SF.
    \item For galaxies with $M_{\rm*} \lesssim 10^{9}\,\rm  M_{\odot}$, \textsc{Shark} predicts a decrease in $q_{\rm IR}$ with decreasing stellar mass and with decreasing redshift. Little observational data exists in that regime, but we note that the likely cause of this in {\sc Shark} is the lack of a model tracking the  relativistic electron escape that is likely to happen in low mass galaxies. 
\end{itemize}

The model for radio continuum emission presented in this paper and implemented in \textsc{Shark} allows the model to extend the wavelength range for SED predictions by orders of magnitude towards the low frequency range, implying an important improvement. We show that this new model is capable of reproducing a variety of observations of galaxies in radio frequencies from the local to the high-redshift Universe RLFs and scaling relations of the radio continuum emission with other galaxy properties. 
Previous literature focusing on predictions for the radio continuum sky have been done through (semi)-empirical models instead \citep{wilman2008semi,bonaldi2019tiered}. Our model contributes to the literature by offering a physical model for the radio continuum emission attached to a physical model of galaxy formation and evolution.

This new model extension provides opportunities to investigate the assumptions that are made with respect to how galaxies are selected in radio continuum relative to the emission in other wavelengths and galaxy properties, and it offers  a tool for future galaxy surveys to predict the expected properties of the galaxies to be observed given certain flux thresholds.

\section*{Data Availability}
The {\sc SURFS} halo and subhalo catalogue and corresponding merger trees used in this work can be accessed from \url{https://tinyurl.com/y6ql46d4}. {\sc Shark}is a public code and the source and python scripts used to produce the plots in this paper can be found at \url{https://github.com/ICRAR/shark/}.

\section*{Acknowledgements}
We thank the anonymous referee for their constructive feedback.

We thank Sabine Bellstedt for sharing observational data that has been used in this work.

CL has received funding from the ARC Centre of
Excellence for All Sky Astrophysics in 3 Dimensions (ASTRO 3D), through project number CE170100013, and is a recipient of the ARC Discovery Project DP210101945. 

LD is a recipient of the ARC Discovery Project FT200100055. 

ID acknowledges support from INAF Minigrant "Harnessing the power of VLBA towards a census of AGN and star formation at high redshift".

This work was supported by resources provided by the Pawsey Supercomputing Centre with funding from the 
Australian Government and the Government of Western Australia.

\bibliographystyle{mnras}
\bibliography{references} 





\bsp	
\label{lastpage}
\end{document}